\documentclass[referee]{raa}

\usepackage{graphicx,times}
\usepackage{orcidlink}  
\usepackage{natbib}
\usepackage{booktabs}
\usepackage{multirow}
\usepackage{amssymb,amsmath}
\usepackage{hyperref}
\bibpunct{(}{)}{;}{a}{}{,}

\begin{document}

  \title{The LAMOST Galaxy Parameter Catalog (LAGC): Stellar Populations, Kinematics, and Emission Lines of Galaxies in the LAMOST Spectroscopic Survey}

   \volnopage{Vol.0 (20xx) No.0, 000--000}
   \setcounter{page}{1}

   \author{Jia-Feng Lu\thanks{E-mail: lujiafeng@westlake.edu.cn} \orcidlink{0000-0002-8817-4587}
      \inst{1}
   \and Shi-Yin Shen\thanks{E-mail: ssy@shao.ac.cn} \orcidlink{0000-0002-3073-5871}
      \inst{3,4}
   \and Qian Liu \orcidlink{0009-0009-0971-5081}
      \inst{2,5}
   \and Shuai Feng \orcidlink{0000-0002-9767-9237}
      \inst{6}
   \and Yong Shi \orcidlink{0000-0002-8614-6275}
      \inst{1}
   }

   \institute{Department of Astronomy, Westlake University, Hangzhou 310030, Zhejiang Province, China; \\
        \and
            Key Laboratory of Optical Astronomy, National Astronomical Observatories, Beijing 100101, China; \\
        \and 
            Shanghai Astronomical Observatory, Chinese Academy of Sciences, No. 80 Nandan Road, Shanghai 200030, People's Republic of China; \\     
        \and 
            Key Lab for Astrophysics, Shanghai, 200034, China; \\
        \and
            University of Chinese Academy of Sciences, Beijing 100049, China;\\
        \and
            College of Physics, Hebei Normal University, 20 South Erhuan Road, Shijiazhuang 050024, China;\\
\vs\no
   {\small Received 20xx month day; accepted 20xx month day}}

\abstract{
The LAMOST Galaxy Parameter Catalog (LAGC) is a value-added data product of the LAMOST spectroscopic survey, generated by an automated spectral-fitting pipeline and released as a continuously updated catalogue series. Despite the vast data volume of the LAMOST spectroscopic survey, no dedicated, homogeneous pipeline had previously been able to extract stellar population, kinematic, and emission-line information simultaneously from its galaxy spectra, primarily because of the poorly controlled absolute flux calibration and a line spread function (LSF) that varies with wavelength, fibre, and time. The pipeline addresses both challenges through the SEW full-spectrum fitting method, which operates in equivalent-width space and is therefore immune to multiplicative flux-calibration errors, together with an accurate, wavelength-dependent model of the LAMOST LSF that enables unbiased kinematic measurements. As the pilot release of the catalogue series, we process the full galaxy sample of the latest LAMOST data release (283,691 spectra) and measure stellar population parameters, stellar and gas kinematics, and equivalent widths of 19 emission lines, each with bootstrap-based uncertainties. Repeat LAMOST observations confirm that the measurements are reproducible and that the bootstrap uncertainties are well calibrated, while comparison with SDSS shows only minor systematic offsets, attributable mainly to the SDSS LSF modelling. The measurements reproduce the well-established relations of the local galaxy population, and the catalogue will be publicly available and continuously updated as new observations become available. LAGC thus provides a reliable scientific catalogue for galaxy evolution studies, while the methodology behind it offers a blueprint for the next generation of fibre-fed spectroscopic surveys.
}

   \authorrunning{J.-F. LU et al.}
   \titlerunning{LAGC}

   \maketitle

\section{Introduction}
\label{sect:intro}

Modern large-scale spectroscopic surveys rely on homogeneous, well-characterised data products to enable robust statistical studies of galaxy formation and evolution. Beginning with the pioneering 2-degree Field Galaxy Redshift Survey \citep[2dFGRS;][]{colless2001} and the Sloan Digital Sky Survey \citep[SDSS;][]{york2000}, and continuing with current facilities such as the Dark Energy Spectroscopic Instrument \citep[DESI;][]{desi2016}, these surveys have mapped the galaxy population across the local Universe in unprecedented detail. A single optical spectrum contains a wealth of information: the stellar continuum and absorption features encode the stellar population history, recoverable through full-spectrum stellar population synthesis \citep{bruzual2003,cidfernandes2005,cappellari2017}; nebular emission lines diagnose the ionising source and trace the recent star formation activity \citep{baldwin1981,kauffmann2003b,kewley2001,brinchmann2004,kennicutt1992}; and the line-of-sight velocity distribution traces the dynamical state of the galaxy \citep{cappellari2013,zahid2016}. These probes have established a pronounced bimodality between young, star-forming and old, quiescent systems \citep{gallazzi2005,strateva2001}, making spectroscopy an essential tool for studying the star formation history, chemical enrichment, and structural evolution of galaxies.

The Large Sky Area Multi-Object Fiber Spectroscopic Telescope \citep[LAMOST;][]{zhao2012,cui2012,luo2012} is a 4\,m-class fibre-fed survey telescope located at the Xinglong Observatory, China. Its unique design combines a large field of view ($5^\circ$ in diameter) with a dense focal plane containing 4,000 optical fibres, enabling the acquisition of up to 4,000 spectra in a single exposure. The low-resolution mode covers 3700--9000~\AA\ at a spectral resolution of $R\sim1800$ \citep{zhao2012}. Although the primary science driver of LAMOST is stellar spectroscopy and Galactic archaeology, the survey also contains a substantial galaxy sample \citep[see][]{shen2016}, complementing SDSS in regions of the northern sky not covered by the latter. As a result, LAMOST has become a valuable resource for galaxy studies based on large statistical samples.

Taking advantage of these spectra, previous studies have exploited the galaxy sample for a range of applications, including galaxy pair identification \citep{shen2016,feng2019}, extreme emission-line compact galaxies \citep{liu2022}, and the galaxy luminosity function \citep{zhao2019}. These studies, however, largely rely on a small number of basic spectral features, the redshift and a few strong lines, for classification and statistical analysis. The full spectrum contains much more information: the continuum and absorption features encode the age and metallicity of the stellar population, while the broadening of absorption and emission lines traces the stellar and gas kinematics. Recovering such physical parameters simultaneously and self-consistently is the key to a far more complete picture of the formation and evolution of LAMOST galaxies.

Obtaining these parameters from LAMOST spectra, however, is not straightforward, and two main challenges must be addressed: the flux-calibration problem and the variable line spread function. The spectra are not absolutely flux calibrated \citep{luo2015,wang2018}; the continuum can deviate systematically from the true spectral energy distribution, with residuals that vary from fibre to fibre and exhibit discontinuities at the junction of the blue and red arms \citep{wang2018,wang2022}. Such multiplicative errors bias full-spectrum stellar population fitting and absorption-line-index measurements \citep[e.g.,][]{bruzual2003,cappellari2004,gallazzi2005}. The variable line spread function (LSF) poses the second of these challenges: it varies with wavelength, fibre, and observation time \citep{smee2013,law2021}, and if not modelled accurately, it biases the measured velocity dispersion and propagates into the stellar population solution through template mismatch. These difficulties are compounded by the well-known degeneracies among stellar populations, kinematics, dust, and flux calibration, which require a fully joint treatment of the parameters \citep{conroy2009,conroy2010,conroy2013,pforr2012,qin2022}.

Previous studies have each recovered one of these quantities from LAMOST spectra: stellar populations \citep{wang2022}, stellar kinematics \citep{napolitano2020}, and emission-line properties \citep{wang2018}. None of these, however, was designed for a joint solution. As a result, despite the large data volume of LAMOST, a homogeneous catalogue that simultaneously provides all three quantities while properly accounting for the variable LSF and the calibration uncertainties has so far been lacking, and no dedicated pipeline has been established to produce such a catalogue in a fully automated, reproducible, and continuously updated manner.

The recently developed SEW method \citep{lu2025sew} provides a physically motivated solution to the continuum-calibration problem. By constructing the spectrum in equivalent-width space, it removes the multiplicative continuum distortions arising from flux calibration and dust attenuation while preserving the narrow absorption and emission features that carry the physical information \citep[e.g.,][]{wilkinson2017,barisic2020,li2020}, without introducing ad hoc additive polynomial corrections or assuming a fixed dust attenuation law \citep[e.g.,][]{fitzpatrick1999,fitzpatrick2007,calzetti2000,cardelli1989}. This treatment obviates the need for per-spectrum user intervention, ensuring a fully automated and reproducible processing of the entire dataset. At the same time, an accurate, wavelength-dependent LSF is essential for unbiased kinematic and stellar population measurements, and for LAMOST this is provided by the machine-learning instrumental profile model of \citet{liu2026}.

Combining these two ingredients, we have built a dedicated processing pipeline that measures stellar population parameters, kinematics, and emission-line equivalent widths simultaneously and automatically. We present this pipeline (LAMOST Galaxy Parameter Pipeline), together with the value-added data products it produces, as the LAMOST Galaxy Parameter Catalog (LAGC). The resulting value-added catalogues are publicly available at the China-VO/NADC data platform, and the catalogue series is continuously updated as new LAMOST observations become available. As the pilot release of the LAGC series, we have processed the complete galaxy sample of the latest LAMOST data release, comprising 283,691 spectra. We validate the output through repeated observations, cross-survey comparison with SDSS, and checks against previous LAMOST results.

This paper is organised as follows. Section~\ref{sect:pipeline} describes the LAMOST Galaxy Parameter Pipeline (LAG2P) in detail: the stellar and gas templates, the line spread function correction, the SEW fitting method, the bootstrap uncertainty estimation, and the resulting value-added data products. Section~\ref{sect:validation} validates the pilot-release value-added catalogue through repeat observations and a cross-survey comparison with SDSS. Section~\ref{sect:results} presents a series of science demonstrations that showcase the scientific capabilities of the catalogue, and Section~\ref{sect:summary} summarises the work and provides recommendations for users. Appendix~\ref{app:examples} presents representative full-spectrum fits of individual galaxies, which illustrate the immunity of the SEW measurements to continuum calibration errors, and Appendix~\ref{app:previous} compares our measurements with previous LAMOST-based catalogues.

\section{The LAMOST Galaxy Parameter Pipeline (LAG2P)}
\label{sect:pipeline}

The LAMOST Galaxy Parameter Pipeline (LAG2P) is the automated spectral-fitting pipeline behind the LAMOST Galaxy Parameter Catalog (LAGC). It measures the stellar population, kinematic, and emission-line parameters of every LAMOST spectrum carrying the galaxy classification \citep{lamostdr13}, regardless of its signal-to-noise ratio or redshift quality, and the resulting value-added data products are publicly available at the China-VO/NADC data platform, and continuously updated as new observations become available. The measurements are based on the SEW full-spectrum fitting method \citep{lu2025sew} and on an accurate, wavelength-dependent model of the LAMOST line spread function.

\begin{figure}
\centering
\includegraphics[width=15cm, angle=0]{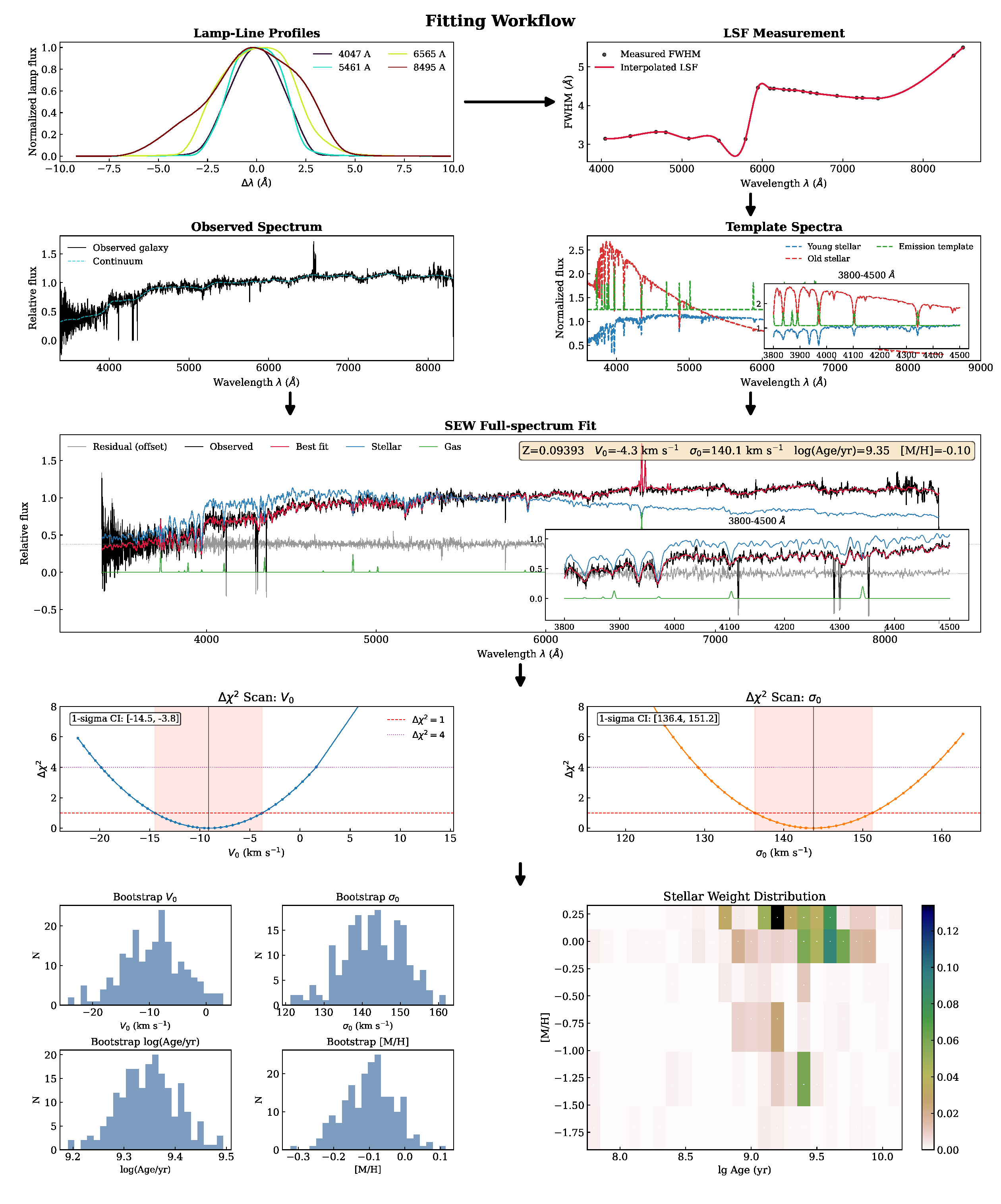}
\caption{Workflow of the LAG2P pipeline. The steps from left to right: (1) de-redshift and mask the input LAMOST spectrum; (2) measure the wavelength-dependent LSF by fitting the machine-learning predicted lamp-line profiles; (3) convolve the SSP templates with the LSF kernel and append emission-line templates; (4) fit the combined model to the EW spectrum with \textsc{pPXF-SEW}, yielding stellar populations, kinematics, and emission-line EWs; (5) bootstrap resampling provides parameter uncertainties.}
\label{fig:pipeline}
\end{figure}

Figure~\ref{fig:pipeline} summarises the processing flow of LAG2P for each LAMOST spectrum. The input spectrum is first de-redshifted to the rest frame, and pixels flagged by the LAMOST data-reduction pipeline (e.g., bad pixels, cosmic-ray hits, and sky residuals) are masked. The LAMOST LSF is then measured from the machine-learning IP model of \citet{liu2026}, which takes the observation time and fibre ID as input and predicts the lamp-line shape. With this LSF, the 144 E-MILES SSP templates are convolved, normalised, and concatenated with the 19 Gaussian emission-line templates. The combined matrix is passed to \textsc{pPXF-SEW}, which simultaneously solves for the six kinematic parameters ($V_\star,\sigma_\star,V_{\text{gas}},\sigma_{\text{gas}},V_{\text{fb}},\sigma_{\text{fb}}$) via non-linear optimisation and for the linear template weights via NNLS. The adaptive bootstrap uncertainty estimation is then applied, and the best-fit gas weights are converted to physical EWs. Finally, light-weighted and mass-weighted stellar population parameters are computed from the SSP weight distribution.

In the following subsections we describe the components of the pipeline that determine the measurements: the stellar and gas templates, the line spread function correction, the SEW fitting engine, and the bootstrap uncertainty estimation.

\subsection{Stellar and gas templates}
\label{sect:templates}

\subsubsection{Stellar population templates}

We adopt the E-MILES simple stellar population (SSP) models \citep{vazdekis2016} as our stellar template library. The E-MILES models are built upon the empirical MILES stellar library \citep{sanchezblazquez2006}, which provides medium-resolution ($R\sim$2000) spectra covering 1,680--50,000~\AA\ for a large set of stars with diverse stellar parameters. The SSP spectra are generated using the Padova isochrones \citep{girardi2000} with a Salpeter initial mass function \citep{salpeter1955}. In our work, the models cover an age range of 0.06--13~Gyr in 24 steps and a metallicity range of [M/H] $= -1.71$ to $+0.22$ in 6 steps, yielding a total of 144 templates.

The E-MILES models have an intrinsic LSF with a constant FWHM of 2.51~\AA  in the range of about 3500--9000~\AA. The correction of the template LSF to the actual LAMOST instrumental LSF is described in Section~\ref{sect:lsf}. All templates are normalised to unity mean flux over the rest-frame wavelength window 5070--5950~\AA.

As the pilot release and demonstration case of the LAG2P catalogue series (Section~\ref{sect:data_products}), the present work is based on the fiducial template set described above as an example. Considering the limitations of this template set, the fixed Salpeter IMF, the solar abundance ratios, and the absence of $\alpha$-enhanced and TP-AGB-rich stellar populations \citep[e.g.,][]{kroupa2001,chabrier2003,thomas2003,marigo2017}, future updates of the catalogue will include parameter measurements derived from a richer set of stellar population templates.

\subsubsection{Emission-line templates}

In addition to the stellar continuum, we construct a set of Gaussian emission-line templates to model nebular emission. The complete set of 19 emission lines is listed in Table~\ref{tab:emission_lines}, along with the velocity component assignments and doublet flux ratios where applicable. Each emission line is modelled as a Gaussian on the logarithmic wavelength grid with an instrumental LSF dispersion. The intrinsic velocity dispersion of each gas component is a free parameter determined by pPXF-SEW.

\begin{table}[htbp]
\centering
\caption{Emission-line set}
\label{tab:emission_lines}
\begin{tabular}{lll}
\toprule
\textbf{Component} & \textbf{Line} & $\lambda_0$ (\AA) \\
\midrule
\multirow{8}{*}{Balmer}
  & H10    & 3798.98 \\
  & H9     & 3836.48 \\
  & H8     & 3890.16 \\
  & H$\epsilon$ & 3971.20 \\
  & H$\delta$  & 4102.90 \\
  & H$\gamma$  & 4341.69 \\
  & H$\beta$   & 4862.69 \\
  & H$\alpha$  & 6564.63 \\
\midrule
\multirow{11}{*}{Forbidden}
  & \relax [OII]$\lambda$3726   & 3727.09 \\
  & \relax [OII]$\lambda$3729   & 3729.88 \\
  & \relax [SII]$\lambda$6716   & 6718.29 \\
  & \relax [SII]$\lambda$6731   & 6732.67 \\
  & \relax [NeIII]$\lambda$3869 & 3869.86 \\
  & \relax [NeIII]$\lambda$3968 & 3968.59 \\
  & HeII$\lambda$4687  & 4687.02 \\
  & HeI$\lambda$5876   & 5877.24 \\
  & \relax [OIII]$\lambda$5007$^\dagger$ & 5008.24 \\
  & \relax [OI]$\lambda$6300$^\ddagger$  & 6302.04 \\
  & \relax [NII]$\lambda$6583$^\star$  & 6585.27 \\
\bottomrule
\multicolumn{3}{l}{$^\dagger$ Tied with [OIII]$\lambda$4960 at ratio 0.33:1.} \\
\multicolumn{3}{l}{$^\ddagger$ Tied with [OI]$\lambda$6365 at ratio 1:0.33.} \\
\multicolumn{3}{l}{$^\star$ Tied with [NII]$\lambda$6550 at ratio 1:0.33.} \\
\end{tabular}
\end{table}

Because the absolute flux calibration of LAMOST spectra is known to be uncertain \citep{luo2015,wang2018}, we do not attempt to measure absolute line fluxes. Instead, we measure the equivalent widths (EWs) of the emission lines, defined as
\begin{equation}
\text{EW} = \int \frac{F_{\text{line}}(\lambda)}{F_{\text{cont}}(\lambda)} d\lambda,
\end{equation}
where $F_{\text{line}}$ is the emission line flux after subtracting the underlying stellar continuum, and $F_{\text{cont}}$ is the continuum level. The EWs are rest-frame equivalent widths in \AA, taken as positive for emission. As we will show below, this quantity is largely invariant to multiplicative flux calibration uncertainties.

Finally, the stellar and emission-line templates are concatenated column-wise into a combined matrix. Kinematically, the three independent components are: the stellar component (all 144 SSP templates), the Balmer emission lines, and the forbidden lines. Each component is parametrised by its own line-of-sight velocity $V$ and velocity dispersion $\sigma$. 

\subsection{Line Spread Function correction}
\label{sect:lsf}

Accurate knowledge of the instrumental line spread function (LSF) is essential for extracting reliable kinematic measurements, particularly the velocity dispersion, and its characterisation is a key step in the data processing of current spectroscopic surveys\citep[e.g.,][]{law2016,guy2023,weilbacher2020}. For LAMOST low-resolution spectra ($R\sim1800$), the LSF varies significantly with wavelength, fibre number, and observation time due to changes in temperature, flexure, and the alignment of the fibre optics.

We characterise the LSF from the machine-learning-based instrumental profile (IP) model of \citet{liu2026}. This model is trained on the full archive of LAMOST arc-lamp exposures, including Hg--Cd lamps for the blue arm (3700--5900~\AA) and Ne--Ar lamps for the red arm (5700--9000~\AA), and compresses the complex, multidimensional variation of the IP into a neural network. Given a query consisting of the observation time, the spectral fibre identifier, and the central wavelength, the network predicts the full IP shape.

For each LAMOST spectrum, we query the machine-learning lamp-spectrum model, fit Gaussian profiles to the predicted IP shapes to obtain the instrumental FWHM at each wavelength, and interpolate the results to produce a continuous function $\text{FWHM}_{\text{gal}}(\lambda)$. When no lamp spectrum is available for the corresponding observation epoch, the nearest available lamp exposure is used instead; the affected spectra are flagged in the catalogue (Section~\ref{sect:data_products}).

With the measured $\text{FWHM}_{\text{gal}}(\lambda)$, we correct the stellar template LSF by convolving each SSP template with a wavelength-dependent Gaussian kernel. Since the E-MILES templates have an intrinsic FWHM of $\text{FWHM}_{\text{tem}} = 2.51$~\AA, the additional broadening required at each wavelength is
\begin{equation}
\sigma_{\rm corr}^2(\lambda) = \frac{\text{FWHM}_{\text{gal}}^2(\lambda) - \text{FWHM}_{\text{tem}}^2}{8\ln 2},
\end{equation}
with a low bound of 0. But in most cases, the LAMOST FWHM is broader than the template.

The same LSF information is used to set the width of the emission-line Gaussian templates.  The intrinsic (astrophysical) velocity dispersion of each gas component is a free parameter determined by pPXF-SEW during the fitting process.

\subsection{Overview of the SEW Method}
\label{sect:SEW_overview}

The SEW method \citep{lu2025sew} is a full-spectrum fitting technique that operates in the equivalent-width (EW) space rather than in the flux space. By replacing the flux at each wavelength with the ratio of the flux to its underlying continuum, the EW spectrum isolates the line and absorption feature information from the continuum shape, which is precisely the part affected by flux calibration uncertainties and dust attenuation. This section summarises the mathematical formulation; details and validation can be found in \citet{lu2025sew}.

\subsubsection{Invariance of the equivalent-width spectrum}

The observed galaxy spectrum $F(\lambda)$ can be expressed as a product of the intrinsic spectrum $F_{\rm intr}(\lambda)$ and a smooth, positive function $P(\lambda)$ that contains both flux calibration residuals and dust attenuation:
\begin{equation}
F(\lambda) = F_{\rm intr}(\lambda) \, P(\lambda).
\end{equation}
The SEW method first constructs the equivalent-width spectrum by taking the ratio of the observed flux to its smoothed continuum $F'(\lambda)$:
\begin{equation}
\text{EW}(\lambda) \equiv \frac{F(\lambda)}{F'(\lambda)},
\label{eq:ew_def}
\end{equation}
where $F'(\lambda)$ is obtained via a robust smoothing procedure (Discrete Penalised Least Squares; see below). Conceptually, this is the pixel-by-pixel analogue of the classical definition of an equivalent width, ${\rm EW} = \int [1 - F(\lambda)/F_c(\lambda)]\,d\lambda$, but applied across the full spectral range.

The key to the invariance is the natural separation of physical scales in galaxy spectra. Stellar absorption features and nebular emission lines are intrinsically narrow in wavelength (typically $\lesssim$\,a few tens~\AA), whereas the continuum shape and the multiplicative factor $P(\lambda)$, whether arising from dust attenuation or flux calibration residuals, vary on much broader scales. This scale separation means that the continuum $F'(\lambda)$ extracted from the observed spectrum retains the same broad-band shape as the data:
\begin{equation}
F'(\lambda) \approx F'_{\rm intr}(\lambda)\,P(\lambda).
\end{equation}
Substituting into equation~\eqref{eq:ew_def}, the $P(\lambda)$ factor cancels:
\begin{equation}
\text{EW}(\lambda) \approx \frac{F_{\rm intr}(\lambda)}{F'_{\rm intr}(\lambda)},
\label{eq:ew_inv}
\end{equation}
which depends only on the intrinsic spectral features and is independent of the flux calibration and dust attenuation.

This invariance is the central advantage of the SEW method. Because the EW spectrum is immune to multiplicative continuum distortions, we can perform the full-spectrum fitting without introducing additive polynomials to absorb calibration residuals or without assuming a specific dust attenuation law a priori.

\subsubsection{Continuum estimation via DPLS}

In the SEW method, the smoothed continuum $F'(\lambda)$ is obtained using the Discrete Penalised Least Squares (DPLS) algorithm \citep{eilers2003}. DPLS finds the continuum that minimises
\begin{equation}
Q = \sum_{i=1}^{n} w_i \bigl(F_i - F'_i\bigr)^2 \;+\; \ell \sum_{i=2}^{n-1} \bigl(\Delta^2 F'_i\bigr)^2,
\label{eq:dpls}
\end{equation}
where $w_i$ are weights, $\Delta^2$ is the second-order difference operator, and $\ell$ is the smoothing parameter. The first term enforces fidelity to the data, while the second penalises curvature in the solution. Because spectral lines and other narrow features have large second derivatives, they are strongly suppressed by the curvature penalty; the broad continuum, which varies slowly and has negligible curvature, is preserved. The smoothing parameter $\ell$ sets the transition scale: features narrower than this scale are treated as curvature and filtered out, while broader variations are retained as part of the continuum.

In this work, we adopt $\ell = 10^{-8}$ for the SEW fitting. As demonstrated in \citet{lu2025sew}, this value corresponds to an effective smoothing window of $\sim$\,0.01 in $\ln\lambda$, or $\sim$\,3000~km\,s$^{-1}$ in velocity space, which robustly separates line features from the continuum. Following their recommendation and to account for the larger flux calibration uncertainties in LAMOST data (e.g., stitching artifacts between the two arms and fibre-to-fibre variations), we choose a relatively smaller $\ell$ within this robust range to let the continuum follow the data more closely, thereby reducing the risk of these calibration artifacts being misinterpreted as intrinsic spectral features. We have verified that varying $\ell$ within the robust range identified by \citet{lu2025sew} does not significantly affect our final results.

The same DPLS procedure is applied to each SSP template $f_j(\lambda)$ to obtain its smoothed continuum $f'_j(\lambda)$. Consequently, the continuum of the full stellar population model follows from the same linear combination of template continua.

\subsubsection{Linearised fitting in EW space}

After obtaining the observed EW spectrum $\text{EW}_{\rm obs}(\lambda)$, we model it as the ratio of the linear combination of SSP templates to the corresponding linear combination of their continua. For a set of template weights $\mathbf{x} = \{x_j\}$, the model EW is
\begin{equation}
\text{EW}_{\rm csp}(\lambda) = \frac{\sum_j x_j f_j(\lambda)}{\sum_j x_j f'_j(\lambda)}.
\label{eq:ew_model}
\end{equation}
This is a non-linear equation in $x_j$ because the denominator depends on the same set of weights. To linearise it, we introduce the normalised weights
\begin{equation}
x'_j = \frac{x_j}{\sum_j x_j}, \qquad \sum_j x'_j = 1.
\end{equation}
Substituting $x_j = x'_j \sum_j x_j$ into Eq.~\eqref{eq:ew_model} and rearranging yields
\begin{equation}
\sum_j \bigl[ f_j(\lambda) - \text{EW}_{\rm csp}(\lambda) \bigl( f'_j(\lambda) - 1 \bigr) \bigr] \, x'_j = \text{EW}_{\rm csp}(\lambda).
\label{eq:linearized}
\end{equation}
For a fixed $\text{EW}_{\rm csp}(\lambda)$ (e.g., set to $\text{EW}_{\rm obs}(\lambda)$ initially), Eq.~\eqref{eq:linearized} is a non-homogeneous linear system in $x'_j$. This avoids the trivial zero solution that would arise from a homogeneous system. The system can be solved under non-negativity constraints using a non-negative least squares (NNLS) solver, and the solution converges within a few iterations because the weights enter only through the slowly varying continuum combination $\sum_j x'_j f'_j(\lambda)$.

The entire fitting procedure is implemented in the public Python package \textsc{pPXF-SEW} \citep{lu2025sew}, an extension of the widely used \textsc{pPXF} code \citep{cappellari2004,cappellari2017}. 

\subsection{Bootstrap uncertainty estimation}
\label{sect:uncertainty}

Full-spectrum fitting involves parameters of two fundamentally different natures. The template weights (stellar population and emission-line amplitudes) enter the model linearly and are solved by NNLS with well-behaved uncertainties. The kinematic parameters, by contrast, are non-linear, and their $\chi^2$ surfaces can be highly non-quadratic, particularly for velocity dispersions near the instrumental resolution limit or for components with weak signal. The usual Hessian-based error estimates, which assume a locally quadratic $\chi^2$ surface, can therefore be unreliable. We adopt a two-phase approach that combines adaptive $\Delta\chi^2$ scanning with bootstrap resampling.

First, for each kinematic parameter separately, we construct an empirical $\Delta\chi^2$ profile while keeping all others fixed. Starting from a coarse grid, we iteratively refine it by inserting new evaluation points at the $\Delta\chi^2 = 1$ and $\Delta\chi^2 = 4$ level crossings and at the midpoints of the largest gaps near the minimum, stopping when the region within $\pm 3\sigma$ is well sampled. The width of each profile provides a robust estimate of the parameter's $1\sigma$ uncertainty without relying on a quadratic approximation.

Using the best-fit model spectrum $F_{\rm fit}(\lambda)$ and these $\sigma$ estimates, we then generate hundreds bootstrap realisations\citep{efron1979}. For each realisation, we draw a new set of kinematic parameters from a multivariate normal distribution, with the velocity dispersions truncated at a lower bound of $\sim$1~km\,s$^{-1}$. A bootstrap spectrum is then constructed by adding Gaussian noise, matching the observed noise, to $F_{\rm fit}$. We re-fit the linear template weights via NNLS for each realisation while holding the drawn kinematics fixed. The mean of the resulting ensemble gives our final estimates for all derived quantities (SSP weights, emission-line strengths, ages, metallicities), and the bootstrap standard deviation provides their associated $1\sigma$ uncertainties. As we demonstrate in Section~\ref{sect:repeat_obs}, the reliability of these error estimates is empirically validated using repeat observations of the same galaxies.

\subsection{Data products and release}
\label{sect:data_products}

The final outputs are stored in a homogeneous catalogue whose contents are summarised in Table~\ref{tab:data_catalog}.

\begin{table}[htbp]
\centering
\caption{Data product catalogue}
\label{tab:data_catalog}
\begin{tabular}{ll}
\toprule
\textbf{Column}  & \textbf{Description} \\
\midrule
\texttt{obsid}  & Observation ID \\
\texttt{ra, dec}  & Right Ascension and Declination (J2000, deg) \\
\texttt{Z} & Redshift \\
\texttt{age, age\_m}  & Luminosity-weighted (at V band) and mass-weighted stellar age (log\,yr) \\
\texttt{metal, metal\_m}  & Luminosity-weighted (at V band) and mass-weighted stellar metallicity [M/H] \\
\texttt{ml\_pop} & Stellar mass-to-light ratio ($M_\odot/L_\odot$) \\
\texttt{sn} & Signal-to-noise ratio in 5 bands ($u,g,r,i,z$) \\
\texttt{star\_weight, star\_weight\_err}  & Weights for the 144 SSP templates and their uncertainties \\
\texttt{gasweight}  & Fitted weights for 19 gas emission-line templates \\
\texttt{params, params\_err}  & Kinematic parameters ($V_{\star},\sigma_{\star},V_{\text{gas}},\sigma_{\text{gas}},V_{\text{fb}},\sigma_{\text{fb}}$) and their uncertainties (km/s) \\
\texttt{age\_err, metal\_err}  & Uncertainties of the light-weighted age and metallicity \\
\texttt{age\_m\_err, metal\_m\_err, ml\_pop\_err}  & Uncertainties of the mass-weighted quantities \\
\texttt{EW\_*}  & Rest-frame equivalent widths of the 19 gas emission lines (\AA)$^\star$ \\
\texttt{EW\_gas\_err}  & Uncertainties of the emission-line equivalent widths \\
\texttt{status}  & Quality flag of the fit (see text) \\
\bottomrule
\multicolumn{2}{p{\textwidth}}{$^\star$ The wildcard `*` denotes the 19 lines listed in Table~\ref{tab:emission_lines}. For the three blends fitted with tied components, the tabulated EW is the sum of both components.} \\
\end{tabular}%
\end{table}

The light-weighted and mass-weighted stellar population parameters, together with the stellar mass-to-light ratio, are all derived from the best-fit SSP weights. The light-weighted quantities weight each template by its $V$-band luminosity. The mass-weighted quantities and the stellar mass-to-light ratio are obtained by further weighting each template by its stellar mass-to-light ratio.

Observational quantities such as the coordinates (\texttt{ra}, \texttt{dec}), the redshift \texttt{Z}, and the signal-to-noise ratios \texttt{sn} are adopted from the LAMOST data-reduction pipeline and are not re-measured by LAG2P; the spectra are de-redshifted with the adopted redshift, and all kinematic velocities are measured relative to it. The uncertainties of the derived parameters come from the bootstrap ensemble described in Section~\ref{sect:uncertainty}: those of the light-weighted stellar population parameters, the SSP weights, and the emission-line equivalent widths are the bootstrap standard deviations, whereas those of the mass-weighted quantities are propagated from the SSP weight uncertainties to first order.

The \texttt{status} column records how each measurement was obtained. A value of \texttt{bootstrap} indicates a successful fit with bootstrap uncertainties; \texttt{noIP} indicates a successful fit for an epoch without lamp-spectrum coverage, for which the nearest available lamp spectrum was substituted (Section~\ref{sect:lsf}) and the uncertainties are therefore estimated analytically rather than by bootstrap; \texttt{fit} indicates that the bootstrap step itself failed, for which analytic uncertainties are likewise reported; and \texttt{fail} indicates that the fitting failed, either for the spectrum itself or because the spectral file is unavailable, in which case all measurements are filled with the sentinel value $-999$. About $0.6\%$ of the galaxy spectra carry the \texttt{fail} flag, mostly spectra dominated by negative flux values that arise from errors in the background subtraction. The flag therefore records the origin of the uncertainties, and does not by itself grade the quality of the measurements. The reliability of the reported uncertainties, and hence the usability of individual measurements, is demonstrated by the repeat-observation error analysis in Section~\ref{sect:repeat_obs}.

The pipeline adopts the standard simplifying assumptions of large spectroscopic surveys: a single line-of-sight velocity distribution for the stellar component, and Gaussian profiles for the nebular emission lines. These choices capture the physics of the majority of galaxies at the resolution and signal-to-noise ratio of the current LAMOST data. Galaxies that require a more precise treatment, such as systems with strong AGN activity or kinematically decoupled stellar populations, are best analysed on a case-by-case basis.

The catalogue series can be obtained from the China-VO/NADC data platform (\url{https://nadc.china-vo.org/res/r101899/}), and is continuously updated as new LAMOST observations become available, with the same pipeline applied to each subsequent release and the resulting measurements appended to the series. 

As the pilot release of the catalogue series, we present the value-added catalogue produced from the latest LAMOST data release, which serves as the demonstration case throughout this work. The catalogue is regenerated in step with the LAMOST data releases and therefore always follows the most recent data; the present version is based on the thirteenth data release (DR13), which encompasses observations from the pilot survey (October 2011) through the thirteenth year survey (May 2025), comprising 6,961 plates and a total of 13.5 million low-resolution spectra. Within this data set, 283,691 spectra are classified as galaxies by the LAMOST pipeline \citep{lamostdr13}. The galaxy sample contains repeated observations, with 41,158 galaxies observed at least twice, yielding 227,239 unique objects; these repeated observations provide a valuable basis for assessing the internal consistency of the pipeline (Section~\ref{sect:validation}).

\begin{figure}
\centering
\includegraphics[width=15cm, angle=0]{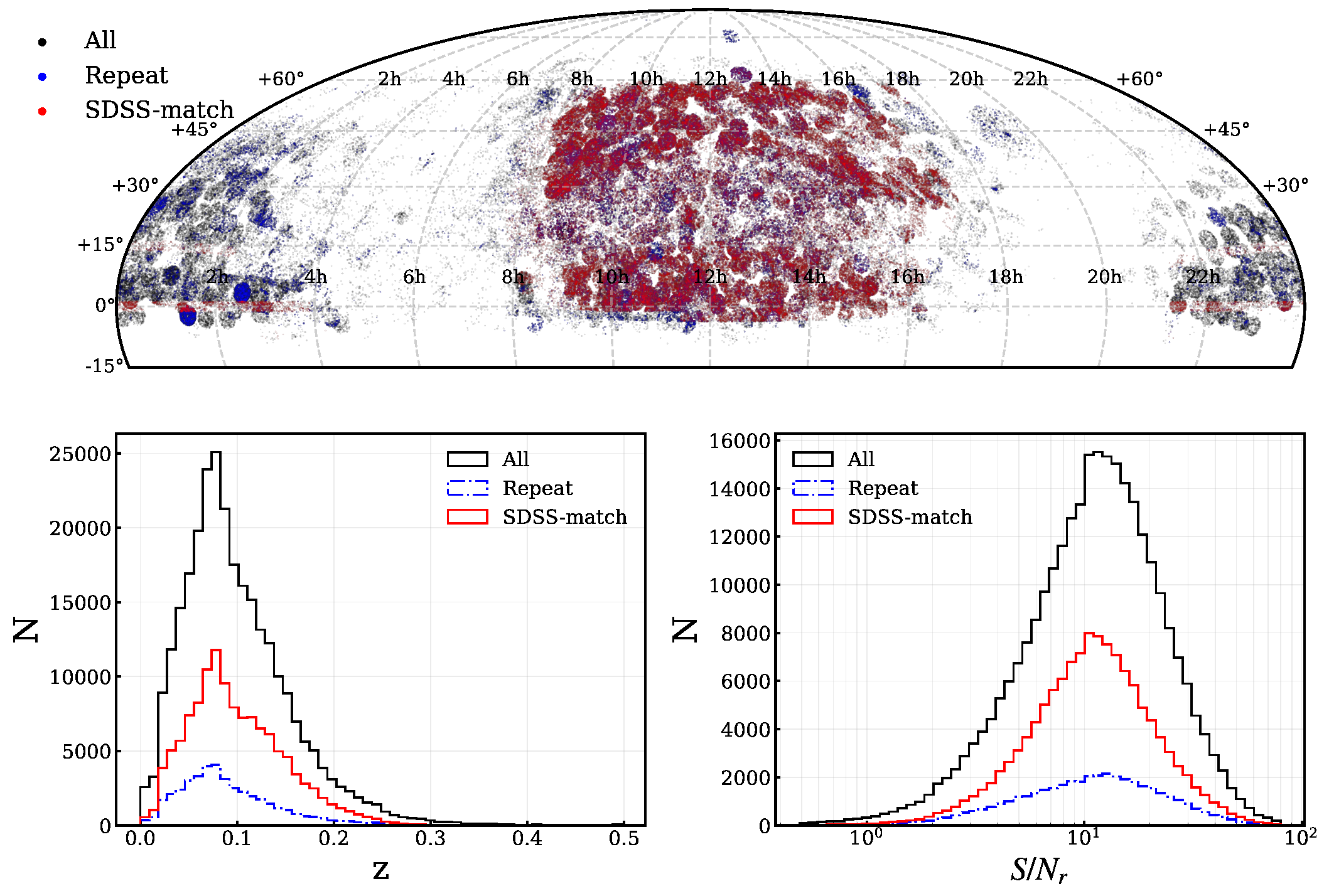}
\caption{Spatial distribution (top panel), redshift distribution (bottom left), and $r$-band signal-to-noise ratio distribution (bottom right) of the LAMOST galaxy sample.}
\label{Lamost_distribution}
\end{figure}

Figure~\ref{Lamost_distribution} shows the spatial distribution, redshift distribution, and signal-to-noise ratio (SNR) distribution of the LAMOST galaxy sample, which covers a wide area of the northern sky across a broad range of redshift and SNR.

\section{Data validation}
\label{sect:validation}

LAG2P is validated on the pilot-release value-added catalogue described in Section~\ref{sect:data_products}. We assess its reliability in two complementary ways: the internal consistency of repeated LAMOST observations (Section~\ref{sect:repeat_obs}) and an external cross-survey comparison with SDSS spectroscopy (Section~\ref{sect:sdss_consistency}). A key advantage of the SEW method is that, by working in the equivalent-width space, the fits are immune to the multiplicative continuum distortions of the LAMOST flux calibration; representative full-spectrum fits of individual galaxies, which illustrate this advantage and the overall quality of the fits, are shown in Appendix~\ref{app:examples}. The SDSS spectra used in the comparison are processed with the same pipeline, adopting a constant instrumental LSF instead of a wavelength-dependent model (Section~\ref{sect:sdss_consistency}). A comparison with previous LAMOST catalogues is given in Appendix~\ref{app:previous}.

\subsection{Repeat observations}
\label{sect:repeat_obs}

\begin{figure}
\centering
\includegraphics[width=15cm, angle=0]{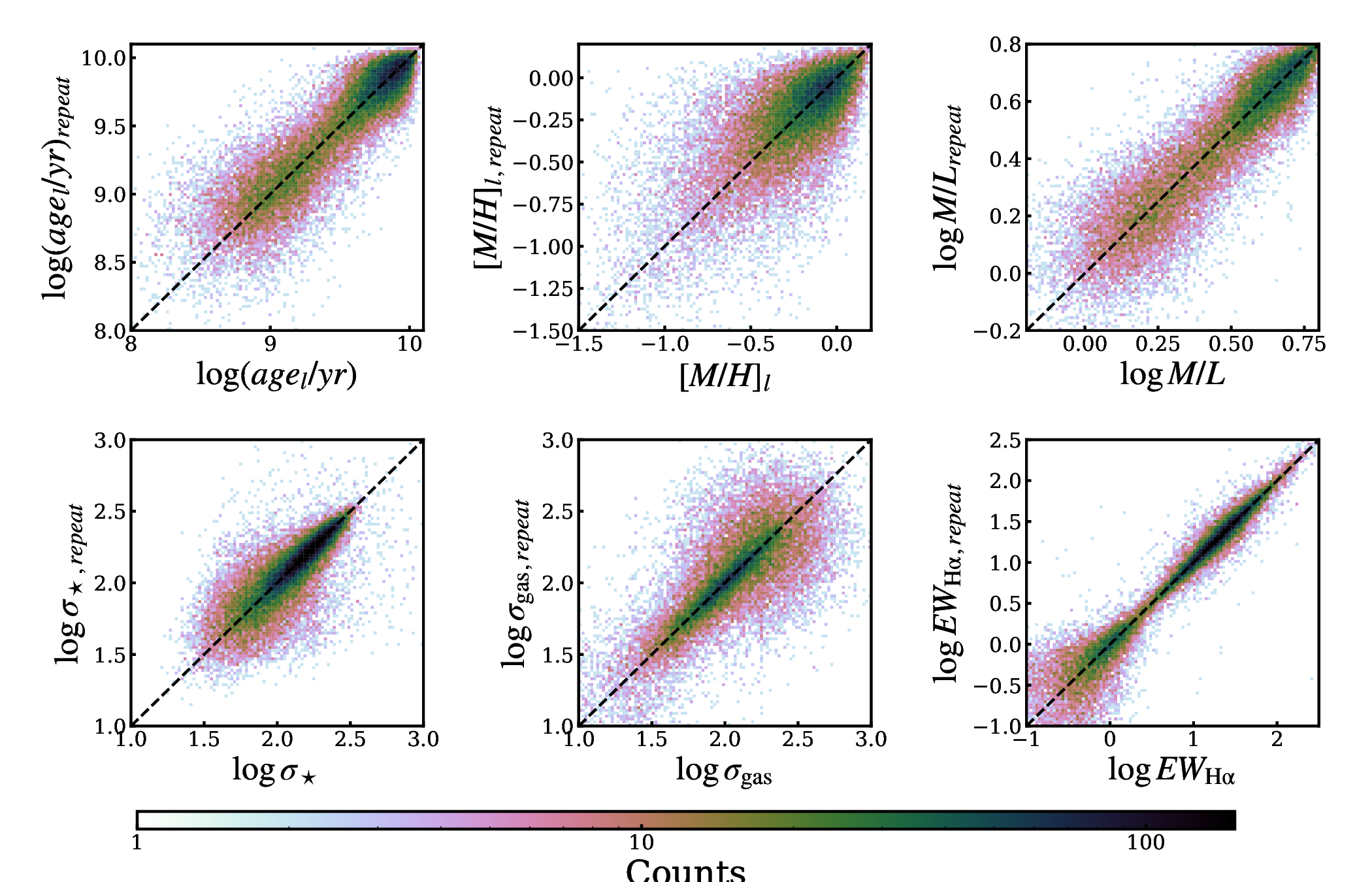}
\caption{Repeat observation comparison of derived parameters between two independent LAMOST exposures of the same galaxies. From left to right and top to bottom: light-weighted age, metallicity, mass-to-light ratio, stellar velocity dispersion, gas velocity dispersion, and H$\alpha$ equivalent width.}
\label{Fig_repeat_compare}
\end{figure}

We assess the internal consistency of the pipeline using the $\sim$\,41,000 galaxies with repeated LAMOST observations; their redshift and SNR distributions closely follow those of the full sample (Figure~\ref{Lamost_distribution}), so this subsample remains representative of the parent sample. Figure~\ref{Fig_repeat_compare} shows the direct comparison of six parameters between two independent exposures of the same galaxies. The measurements are tightly clustered around the one-to-one relation across the full dynamic range, demonstrating excellent reproducibility. At the level of individual objects, the two LAMOST exposures of the representative galaxies in Appendix~\ref{app:examples} produce nearly identical full-spectrum fits and overlapping equivalent-width spectra, confirming that the SEW fits are insensitive to the calibration differences between the two epochs.

\begin{figure}
\centering
\includegraphics[width=15cm, angle=0]{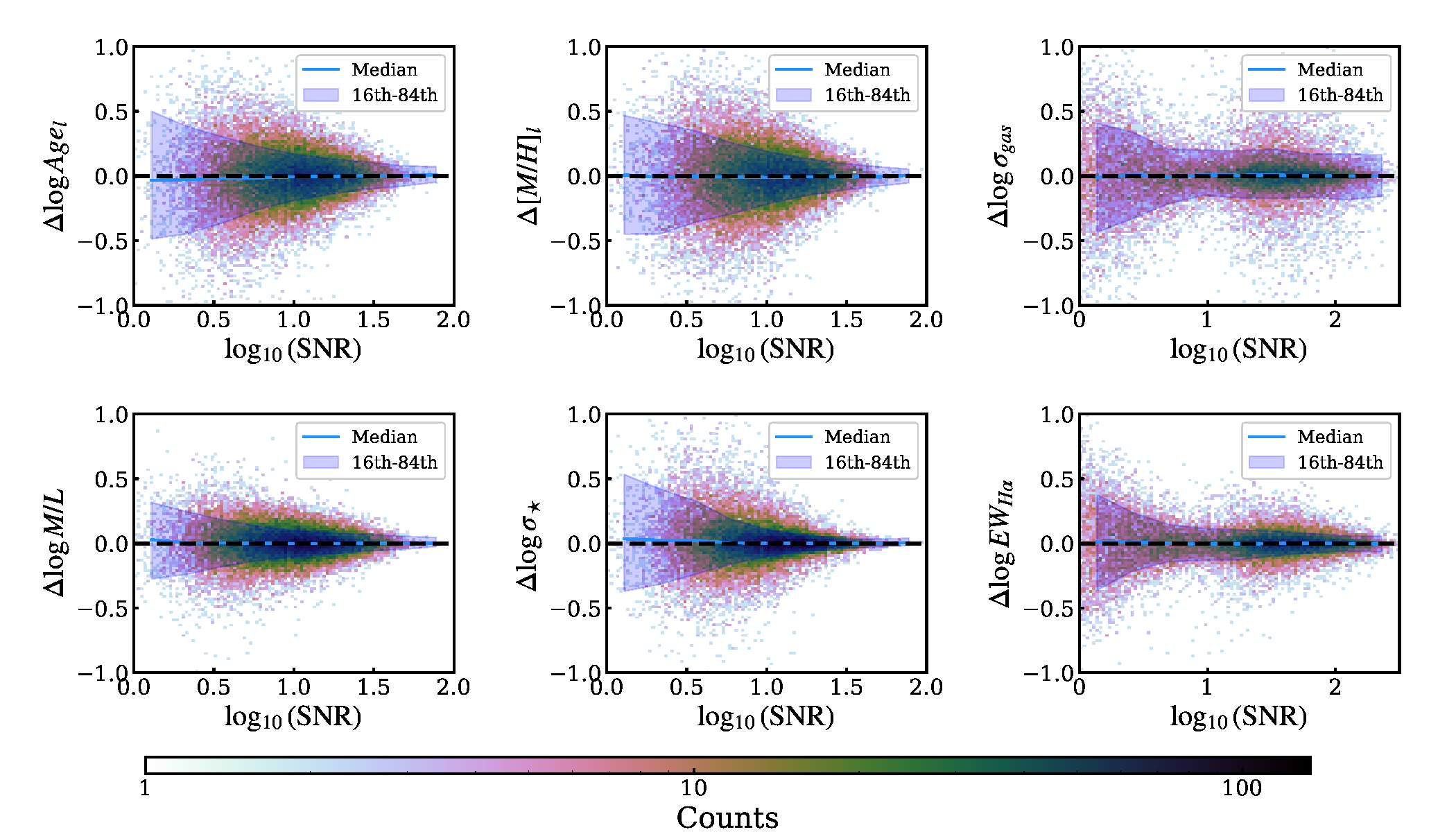}
\caption{LAMOST repeat observation deviations as a function of SNR for six parameters: $\Delta\log\text{Age}_l$, $\Delta[\text{M/H}]_l$, $\Delta\log\sigma_{\star}$, $\Delta\log\sigma_{\text{gas}}$, $\Delta\log M/L$, and $\Delta\log\text{EW}_{\text{H}\alpha}$. The blue line and shaded region show the median and 16th--84th percentile ranges.}
\label{Fig_repeat_deviation}
\end{figure}

The deviations between repeat measurements are further examined as a function of SNR in Figure~\ref{Fig_repeat_deviation}. For all six parameters, the median deviation is consistent with zero at all SNR levels, confirming the absence of systematic biases. The 16th--84th percentile scatter decreases with increasing SNR as expected, indicating that the measurements are limited by photon statistics rather than systematics.

\begin{figure}
\centering
\includegraphics[width=15cm, angle=0]{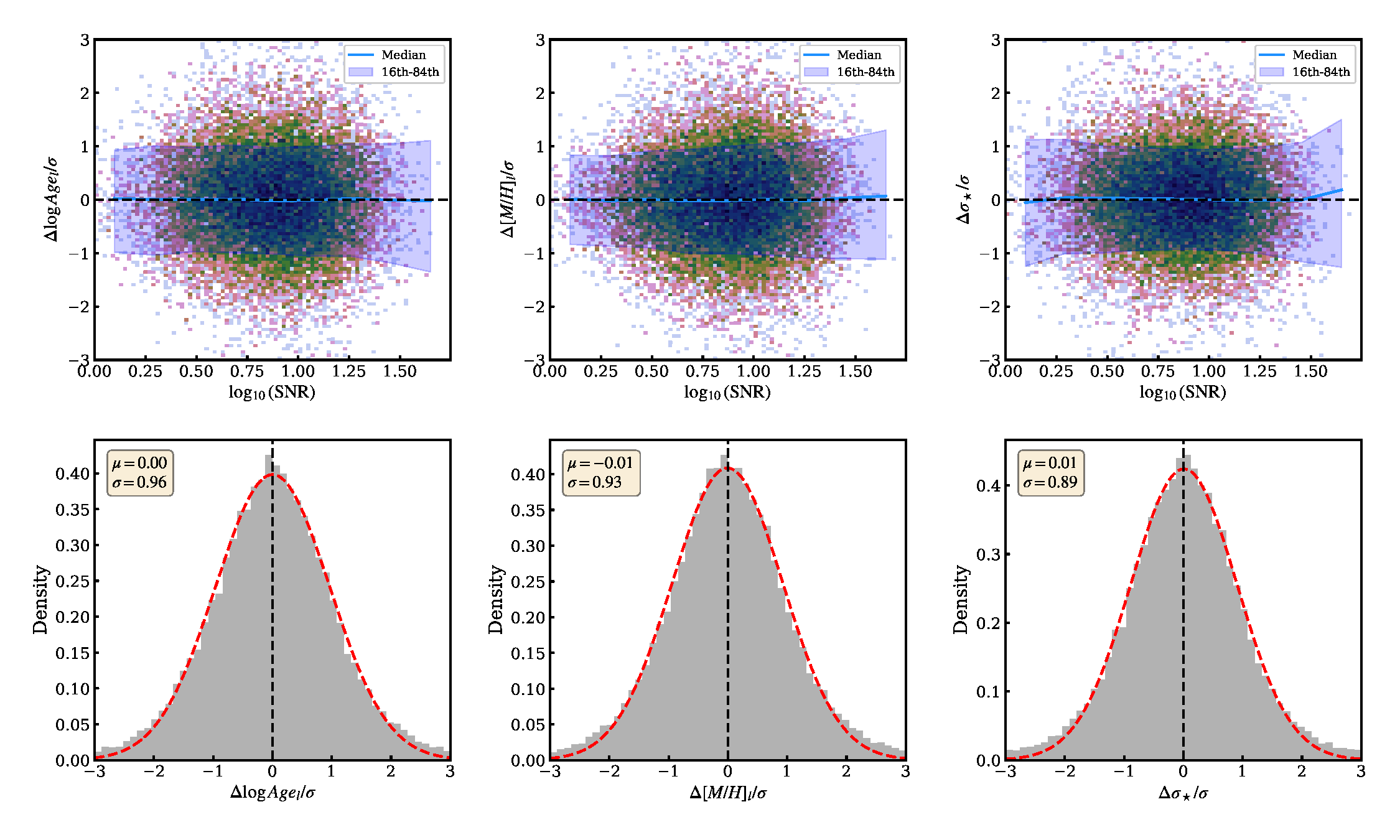}
\caption{Validation of bootstrap uncertainty estimates using LAMOST repeat observations. Top panels: normalised deviations of light-weighted age (left), metallicity (middle), and stellar velocity dispersion (right) as a function of SNR. The blue line and shaded region show the median and 16th--84th percentile ranges. Bottom panels: marginal distributions of the normalised deviations with Gaussian fits (red dashed curves). The fit widths of $\sigma \sim 1$ confirm that the bootstrap errors are well calibrated.}
\label{Fig_noise}
\end{figure}

The bootstrap uncertainty estimation (Section~\ref{sect:uncertainty}) accounts for the non-linear nature of the kinematic parameters and the correlations between different fitted components, which is particularly important for the interpretation of scaling relations and statistical analyses. We validate these error estimates empirically using the same repeat-observation sample. For each pair of repeat observations, we compute the normalised deviation $(\theta_1 - \theta_2) / \sqrt{\sigma_1^2 + \sigma_2^2}$, where $\sigma_1$ and $\sigma_2$ are the bootstrap uncertainties from the two independent fits. If the error estimates are well calibrated, these normalised deviations should follow a standard normal distribution with unit variance. Figure~\ref{Fig_noise} examines these normalised deviations for three representative parameters: light-weighted age, metallicity, and stellar velocity dispersion. The top panels show the deviations as a function of SNR; the median is consistent with zero across the full SNR range, and the 16th--84th percentile width remains approximately $\pm 1$ at all SNR levels. The bottom panels show the marginal distributions, which are well described by Gaussian fits with $\sigma \approx 1$, confirming that the bootstrap uncertainties provide a faithful representation of the true measurement errors without systematic bias.

\subsection{External consistency with SDSS}
\label{sect:sdss_consistency}

\begin{figure}
\centering
\includegraphics[width=15cm, angle=0]{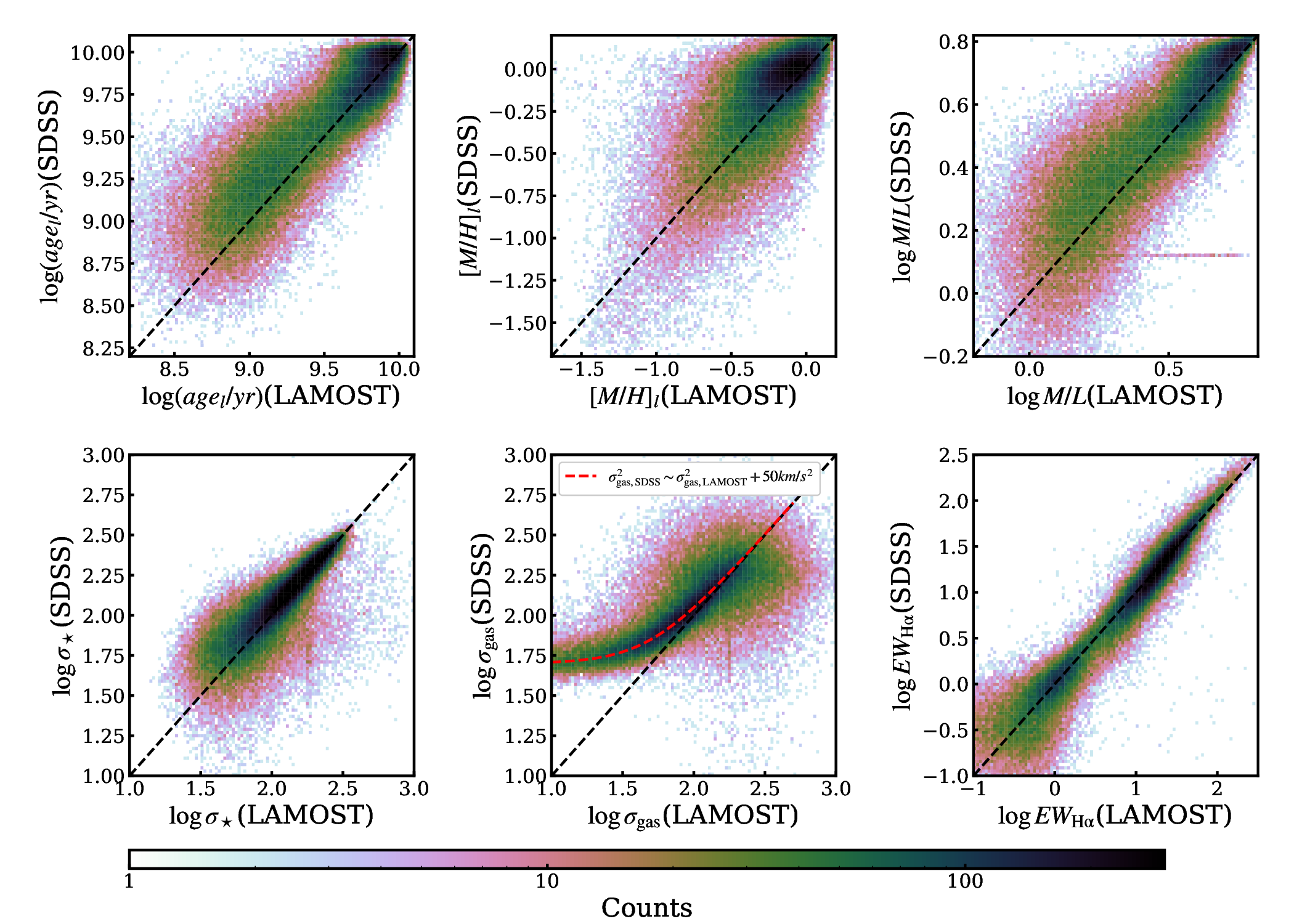}
\caption{Comparison of derived parameters between LAMOST and SDSS for galaxies with matching observations. From left to right: light-weighted age, metallicity, stellar velocity dispersion, gas velocity dispersion, mass-to-light ratio, and H$\alpha$ equivalent width. The dashed line shows the one-to-one relation.}
\label{Fig6}
\end{figure}

For external validation, we match the LAMOST sample with SDSS DR7 \citep{abazajian2009} spectroscopic galaxies using a positional tolerance of 1\,arcsec, yielding 125,388 galaxies with spectra from both surveys; their redshift and SNR distributions closely match those of the full sample (red points in Figure~\ref{Lamost_distribution}), confirming that the matched subsample remains representative of the parent sample. For each matched galaxy we independently run the same pipeline on the SDSS spectrum, adopting a constant FWHM of 2.76~\AA\ for the instrumental LSF, corresponding to the typical resolving power of $R\sim2000$ at 5500~\AA\ ($R = \lambda/\Delta\lambda$); a wavelength-dependent LSF cannot be derived for SDSS because no complete set of arc-lamp exposures is available. Figure~\ref{Fig6} shows the direct comparison of six key parameters: light-weighted age, metallicity, stellar velocity dispersion, gas velocity dispersion, mass-to-light ratio, and H$\alpha$ EW. As illustrated by the representative galaxies in Appendix~\ref{app:examples}, although the observed LAMOST and SDSS continua differ, the equivalent-width spectra agree within the errors, so the cross-survey comparison directly tests the calibration-invariant SEW measurements.

Overall, SDSS measurements are systematically slightly higher than LAMOST for both the stellar population parameters and the kinematics. The median offsets are $\Delta\log\text{Age} \approx -0.08$\,dex, $\Delta[\text{M/H}] \approx -0.08$\,dex, and $\Delta\log\sigma_{\star} \approx -0.02$\,dex ($\sim$\,$-4\%$). The corresponding scatters are $\sim$\,0.21\,dex, $\sim$\,0.21\,dex, and $\sim$\,0.11\,dex. Considering the various systematic factors discussed below, these offsets are within the typical systematics expected for such cross-instrument comparisons. The gas velocity dispersions show a systematic relation, $\sigma_{\text{gas,SDSS}}^2 \approx \sigma_{\text{gas,LAMOST}}^2 + (50~\text{km\,s}^{-1})^2$ (Figure~\ref{Fig6}), which we attribute to the combined effect of the different fibre apertures and the simplified LSF model used for the SDSS spectra. This empirical relation provides a useful calibration for users combining the two surveys.

\begin{figure}
\centering
\includegraphics[width=15cm, angle=0]{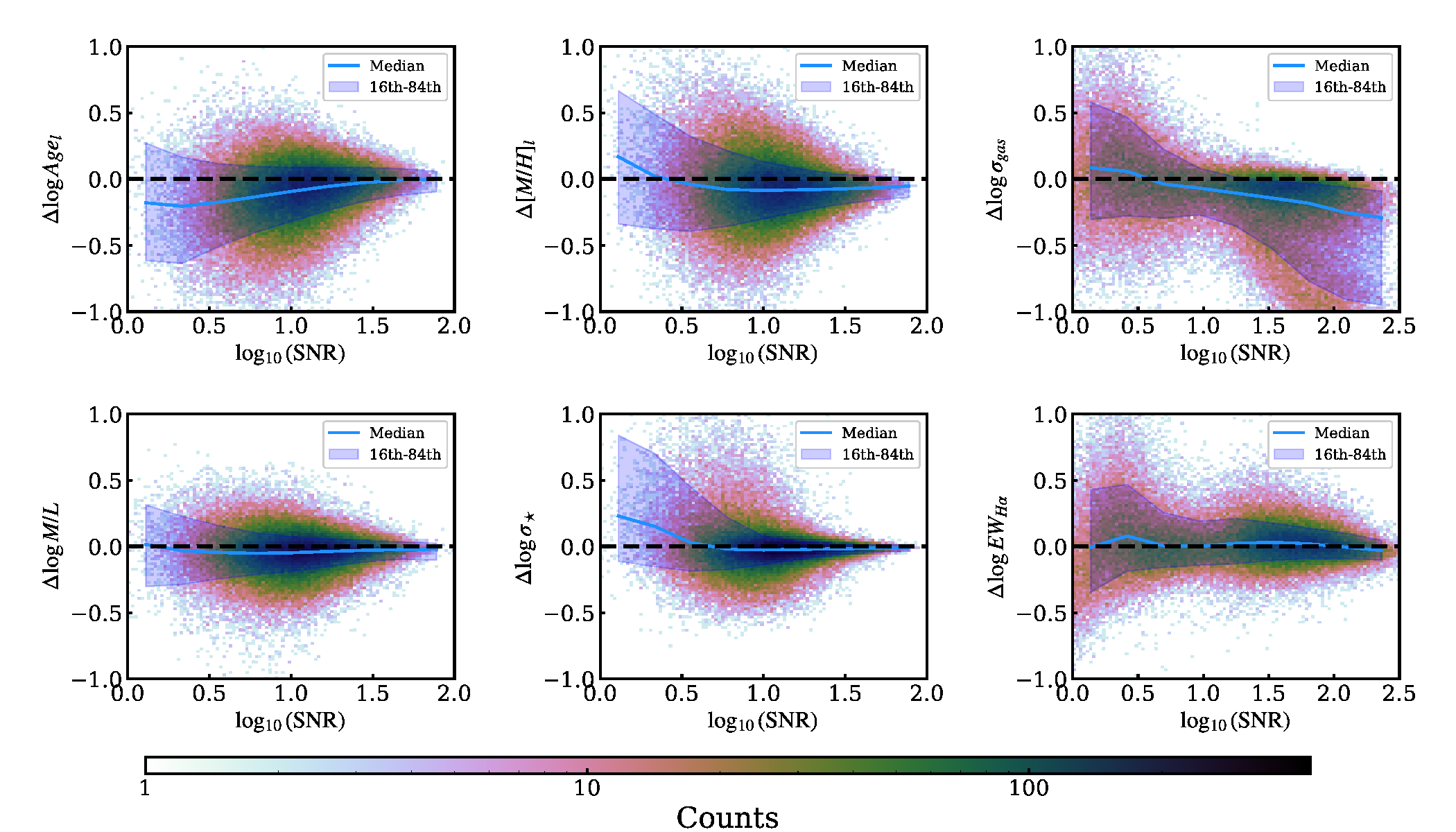}
\caption{LAMOST--SDSS parameter deviations as a function of SNR. The blue line and shaded region show the median and 16th--84th percentile ranges. The horizontal dashed line marks zero deviation.}
\label{Fig7}
\end{figure}

The SNR dependence of these deviations is shown in Figure~\ref{Fig7}. For the stellar population parameters and $\sigma_{\star}$, the scatter decreases systematically with increasing SNR and the median offsets converge toward zero. The $1\sigma$ scatter declines from $\sim$\,0.43 to $\sim$\,0.08\,dex in $\log\text{Age}$, from $\sim$\,0.49 to $\sim$\,0.08\,dex in $[\text{M/H}]$, and from $\sim$\,0.46 to $\sim$\,0.03\,dex in $\sigma_{\star}$ over the same SNR range. In contrast, the $\sigma_{\text{gas}}$ difference does not converge toward zero at high SNR ($\Delta\log\sigma_{\text{gas}} \approx -0.20$\,dex at SNR $> 56$), confirming that it reflects a systematic, rather than stochastic, difference between the two surveys.

These systematic differences are likely caused by a combination of factors, most notably the aperture mismatch between the two surveys and the simplified LSF model adopted for the SDSS spectra. The aperture mismatch (LAMOST 3.3\,arcsec vs.\ SDSS 3\,arcsec, with LAMOST's effective diameter varying from 3.6 to 4.9\,m) leads to differences in the sampled physical regions, particularly for galaxies with strong radial gradients \citep{zhu2023}. In addition, the SDSS spectra are fitted with a constant FWHM of 2.76~\AA, whereas the true SDSS LSF varies with wavelength. Such an LSF mismatch may bias the inferred stellar populations and kinematics, and is most relevant for the gas velocity dispersion, which is constrained mainly by H$\alpha$, where the assumed constant FWHM is too narrow compared to the true SDSS LSF of $\sim$3.7~\AA. The comparison thus highlights that a wavelength-dependent characterisation of the instrumental LSF, as implemented in our LAMOST fitting through the IP model of \citet{liu2026}, is important for unbiased kinematic measurements, especially for dispersion-dominated quantities such as $\sigma_{\text{gas}}$.

The systematic comparison of our measurements with previous LAMOST-based catalogues of stellar velocity dispersions, emission-line ratios, and stellar populations is presented in Appendix~\ref{app:previous}.

\section{Science demonstrations}
\label{sect:results}

To demonstrate the scientific utility of the value-added catalogue, we present in this section a series of example applications selected to exercise each of the main measurements: stellar populations (Section~\ref{sect:stellarpop}), emission-line equivalent widths (Section~\ref{sect:emission_results}), and kinematics (Section~\ref{sect:kinematic_results}). These demonstrations show that the catalogue reproduces well-established relations of the local galaxy population and, in several cases, already reveals results that go beyond previous work. They therefore serve both as showcases of the scientific capabilities of the catalogue and as a science-level validation of the pipeline; a full scientific exploitation of the sample is left to dedicated follow-up studies.

\subsection{Stellar population results}
\label{sect:stellarpop}

We first inspect the stellar populations of LAMOST galaxies, using the light-weighted and mass-weighted ages and metallicities provided by the catalogue (Section~\ref{sect:data_products}). Because the integrated light of a galaxy is dominated by its youngest, most luminous stellar populations, the light-weighted quantities are strongly influenced by recent star formation, whereas the mass-weighted quantities are weighted towards the older stellar populations that carry most of the stellar mass. The difference between the two therefore encodes the star formation and metal enrichment history of a galaxy.

To examine how the stellar populations depend on galaxy mass and star formation activity, we bin our sample by stellar mass and specific star formation rate (sSFR) \citep{brinchmann2004,noeske2007,speagle2014}, adopting stellar masses and star formation rates from the MPA-JHU value-added catalogue of SDSS DR7 \citep{kauffmann2003a,brinchmann2004}. Galaxies with $\log\mathrm{sSFR} > -11$ are classified as star-forming, and those with $\log\mathrm{sSFR} < -11$ as quiescent.

\begin{figure}
\centering
\includegraphics[width=15cm, angle=0]{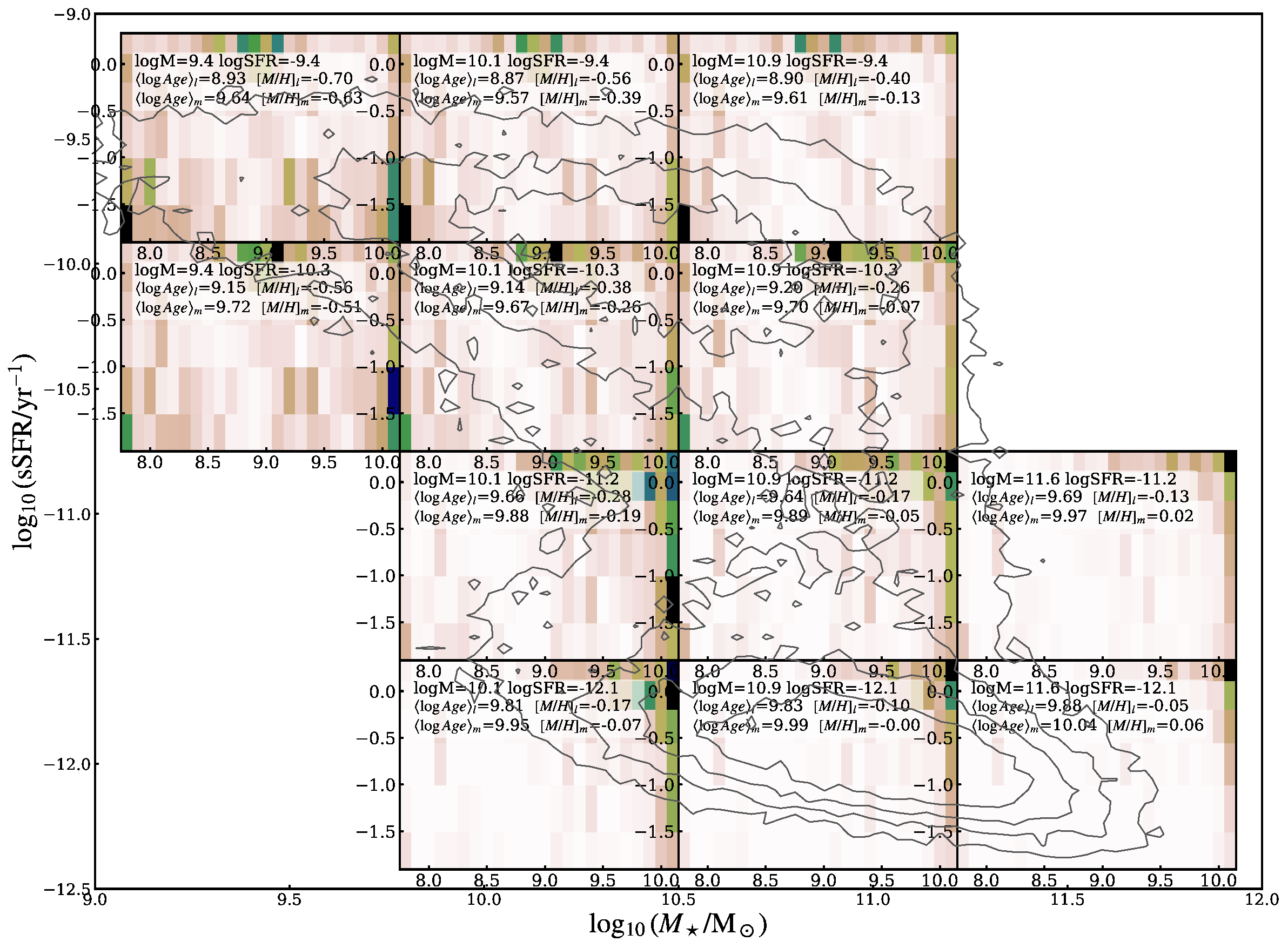}
\caption{Stacked stellar population distributions of LAMOST galaxies on the star-forming main sequence. Each sub-panel shows the distribution of the fitted SSP weights in the age--metallicity plane for galaxies in the corresponding stellar mass (abscissa) and sSFR (ordinate) bin, with the background contours tracing the star-forming main sequence of the full sample. The labels give the median light-weighted and mass-weighted ages and metallicities of each bin.}
\label{Fig_sfms}
\end{figure}

Figure~\ref{Fig_sfms} shows the stacked SSP weight distributions in the age--metallicity plane for galaxies in different mass and sSFR bins. At fixed sSFR, the old stellar populations of low-mass galaxies are more metal-poor than those of high-mass galaxies, following the stellar mass--metallicity relation \citep{gallazzi2005,Panter2008}: for the most quiescent galaxies, for example, the median metallicity of the old stellar populations rises from [M/H]$\sim-0.2$ to $\sim0$ as $\log M_\star$ increases from $\sim10$ to $\sim11.6$. At fixed stellar mass, the distributions become more extended and diffuse with increasing sSFR, with a larger contribution from young, metal-poor stellar populations that reflects ongoing star formation; accordingly, the light-weighted age decreases from $\log\text{Age}\sim9.8$ in the most quiescent bins to $\sim8.9$ in the most star-forming bins, and the light-weighted metallicity drops by $\sim0.3$--$0.4$\,dex. Quiescent galaxies \citep[e.g.,][]{balogh1999}, whose star formation ceased early in their history, are dominated by old stellar populations, with a metallicity that increases towards higher masses.

\begin{figure}
\centering
\includegraphics[width=15cm, angle=0]{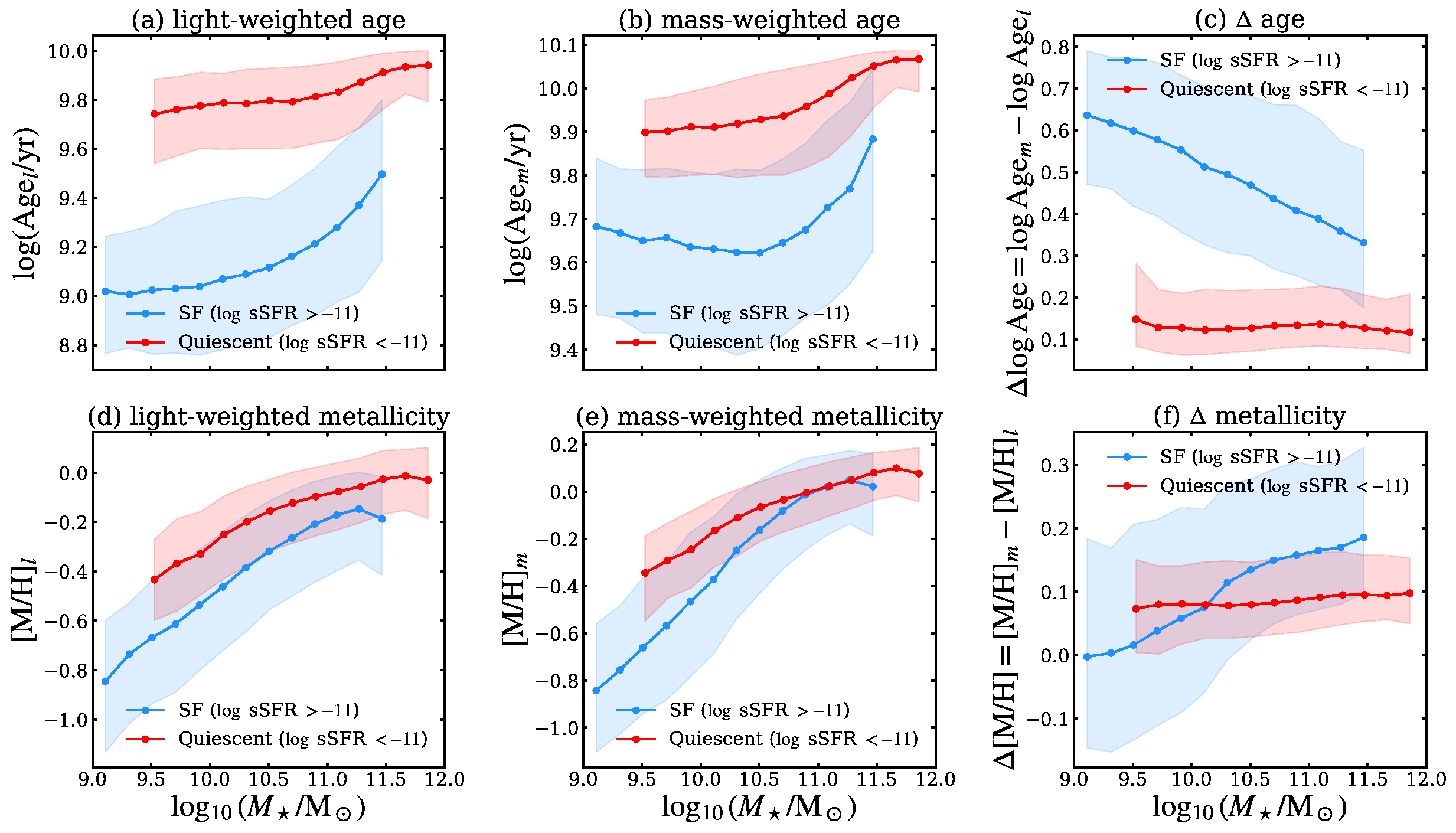}
\caption{Stellar ages (top) and metallicities (bottom) as functions of stellar mass. Left and middle columns show the light-weighted and mass-weighted quantities, respectively, and the right column shows their differences. Blue and red curves are the binned medians (with 16th--84th percentile ranges shaded) for star-forming ($\log\,\mathrm{sSFR}>-11$) and quiescent ($\log\,\mathrm{sSFR}<-11$) galaxies.}
\label{Fig_massrel}
\end{figure}

Figure~\ref{Fig_massrel} quantifies these trends, showing the light-weighted and mass-weighted ages and metallicities, and their differences, as functions of stellar mass.

For quiescent galaxies ($\log\mathrm{sSFR}<-11$), both the age and the metallicity increase with stellar mass, while the light-/mass-weighted differences remain nearly constant, with $\Delta\log\text{Age}\approx0.1$--$0.15$\,dex and $\Delta[\text{M/H}]\approx0.1$\,dex across the full mass range. This is expected for systems whose star formation ceased early: their stellar populations evolve passively, without recent star formation or fresh gas inflows, so the light-weighted and mass-weighted quantities remain tied to the same old stellar component, and the differences stay constant.

Star-forming galaxies also become older and more metal-rich with increasing mass, but they show a systematic light-/mass-weighted offset whose magnitude varies with mass. The age difference $\Delta\log\text{Age}$ decreases from $\sim0.65$\,dex at low mass to $\sim0.35$\,dex at high mass. Because the light-weighted age is biased towards the recently formed stellar component, this difference directly traces the strength of recent star formation and behaves analogously to the sSFR; its decline with mass therefore reflects the well-known downsizing trend \citep{cowie1996,thomas2010}, in which more massive galaxies completed a larger fraction of their star formation at earlier epochs. The metallicity difference $\Delta[\text{M/H}]$, in contrast, increases with mass, from $\sim0$ at low mass to $\sim0.2$\,dex at high mass. The mass-weighted metallicity is dominated by the old, long-enriched stellar population, whereas the light-weighted value is pulled down by young stars. A plausible interpretation of the growing difference is that more massive star-forming galaxies sustain their star formation through stronger, ongoing gas inflows, which dilute the metallicity of the newly formed stars \citep{cresci2010,maiolino2019,sanchezalmeida2014}. Such a short-timescale process would be superimposed on the long-term mass--metallicity enrichment history encoded in the overall increase of the stellar metallicity with mass. Previous studies of gas inflow and its metal-diluting effect have approached the problem from the gas phase, tracing the accretion directly in the ionised-gas abundances of individual galaxies. Our stellar-side measurement is compatible with this picture: the systematically growing light-/mass-weighted metallicity difference with stellar mass would then carry the imprint of the inflowing gas on the newly formed stars, connecting the two sides of the same process on a population level. Model-related systematics, such as the treatment of the thermally pulsing AGB phase \citep[e.g.,][]{marigo2017} and the residual age--metallicity degeneracy, may also contribute to the trend. Assessing their impact will require the richer stellar population templates planned for future releases (Section~\ref{sect:templates}).

\subsection{Emission line results}
\label{sect:emission_results}

\begin{figure}
\centering
\includegraphics[width=15cm, angle=0]{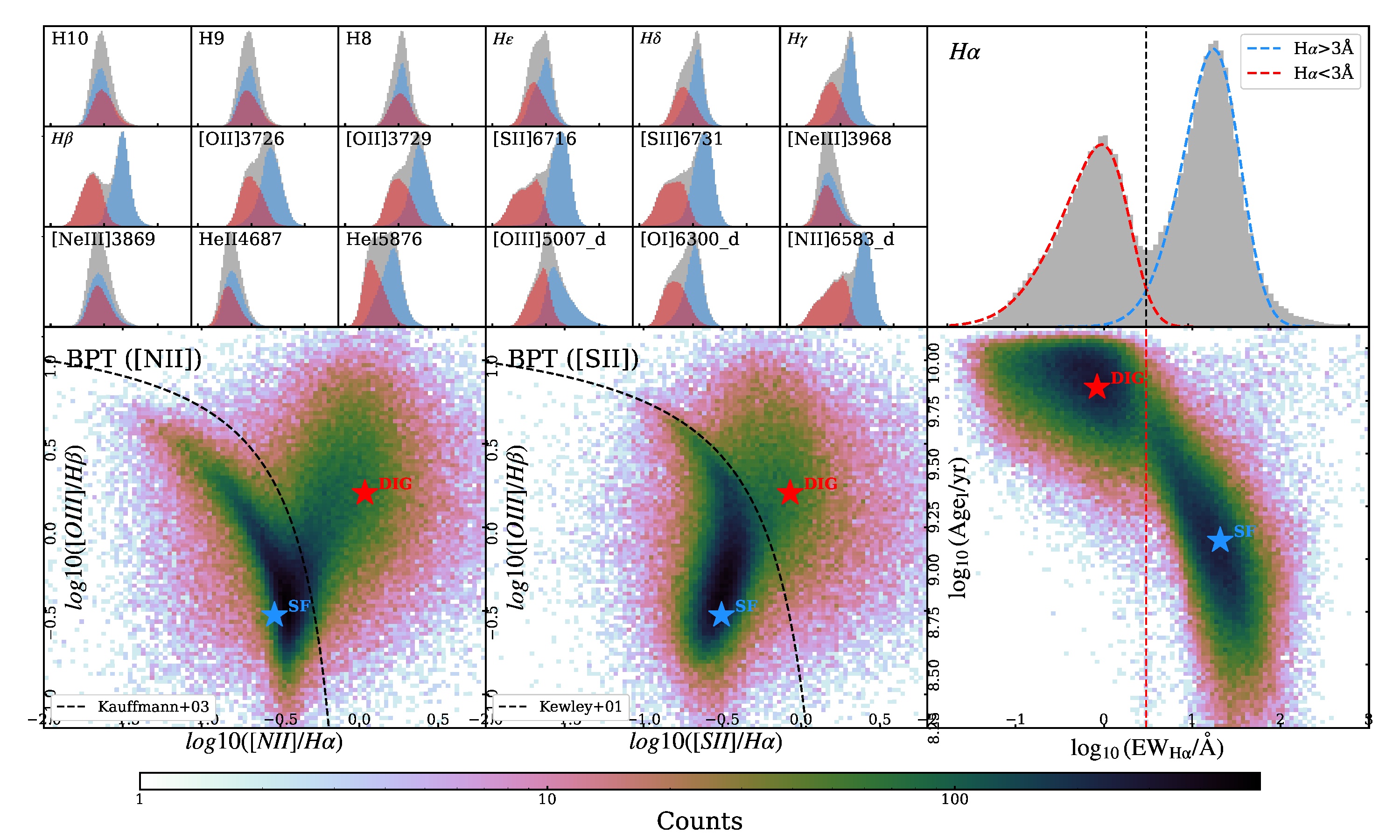}
\caption{Emission-line equivalent width measurements. Top and side panels: distributions of the EWs of the 19 emission lines. For the 18 lines other than H$\alpha$, the grey, blue and red histograms show the full sample and the H$\alpha>3$~\AA\ and H$\alpha<3$~\AA\ subsamples, respectively. The H$\alpha$ panel shows the bimodal distribution fitted with two skew-normal components (dashed curves), with the vertical dashed line marking the 3~\AA\ division. Bottom: BPT diagrams based on EW ratios, with the dashed lines showing the \citet{kauffmann2003b} classification boundary (left, $[\text{NII}]$ BPT) and the \citet{kewley2001} extreme starburst line (middle, $[\text{SII}]$ BPT). Bottom right: EW$_{\text{H}\alpha}$ versus light-weighted age, showing the strong anti-correlation. The blue and red star symbols mark the locations of the high-EW (H$\alpha>3$) and low-EW (H$\alpha<3$) characteristic line sets of Table~\ref{tab:emission_peaks}: the former falls in the star-forming region with young ages, the latter in the low-ionisation region with old ages.}
\label{Fig_emmission}
\end{figure}

Complementing this, we explore the emission-line properties of the sample through the equivalent widths (EWs) stored in the catalogue. Figure~\ref{Fig_emmission} shows the distributions of the EWs of the 19 fitted emission lines, together with the BPT diagnostic diagrams constructed from EW ratios and the EW$_{\text{H}\alpha}$--age relation. Among the Balmer lines, H$\alpha$ is by far the strongest, followed by H$\beta$ and the higher-order Balmer lines H$\gamma$, H$\delta$, and so on; the strongest forbidden lines are the [OII]$\lambda\lambda$3726,3729 doublet and [NII]$\lambda$6583, followed by [SII]$\lambda\lambda$6716,6731 and [OIII]$\lambda$5007, and the weaker [NeIII], [OI], HeI and HeII lines.

The BPT diagrams \citep{baldwin1981} constructed from our EW ratios are shown in the bottom panels of Figure~\ref{Fig_emmission}. The vast majority of the galaxies lie in the star-forming region shown by the \citet{kauffmann2003b} and \citet{kewley2001} classification boundaries, and the overall morphology of both the [NII] and [SII] diagrams closely resembles that found for SDSS galaxies. The bottom-right panel shows that EW$_{\text{H}\alpha}$ is strongly anti-correlated with the light-weighted stellar age: the median age decreases monotonically with increasing EW$_{\text{H}\alpha}$, from $\log\text{Age}\approx9.8$ at the lowest EWs to $\log\text{Age}\approx8.8$ at the highest. In other words, galaxies with larger emission-line equivalent widths are systematically younger, confirming that the SEW-based EWs provide a reliable, flux-calibration-independent tracer of the recent star formation activity.

Because EW$_{\text{H}\alpha}$ is such a direct tracer of the recent star formation activity and stellar age, its distribution naturally encodes the division between the two galaxy populations. Indeed, the H$\alpha$ EW distribution (top-right panel of Figure~\ref{Fig_emmission}) spans more than two orders of magnitude and is clearly bimodal, with a deep valley at EW$\sim$3~\AA\ separating a peak of quiescent galaxies at EW$\sim$0.9~\AA\ from a peak of star-forming galaxies at EW$\sim$20~\AA. Roughly 46\% of the galaxies with detected H$\alpha$ have EW$<3$~\AA\ and $\sim$54\% have EW$>3$~\AA. A gap at EW$_{\text{H}\alpha}\sim3$~\AA\ has been reported by previous large spectroscopic surveys \citep{kauffmann2003b,kewley2006,lacerda2018,lacerda2020}, and is commonly interpreted as the division between galaxies whose nebular emission is dominated by star-forming HII regions and those dominated by diffuse ionised gas (DIG) \citep{lacerda2018,lacerda2020}. This motivates our adoption of EW$_{\text{H}\alpha}=3$~\AA\ to divide quiescent from star-forming galaxies throughout this work \citep{cidfernandes2011}.

The bimodality is not unique to H$\alpha$. When the sample is split by the H$\alpha$\,3~\AA\ criterion, most of the other strong lines also show a clear two-component structure. Their EW distributions are shifted towards high values for H$\alpha>3$~\AA\ (star-forming) galaxies and towards low values for H$\alpha<3$~\AA\ (quiescent) galaxies (blue and red histograms in Figure~\ref{Fig_emmission}). Table~\ref{tab:emission_peaks} lists the peak positions of the two components for each line. The two sets of characteristic line strengths, the high-EW set and the low-EW set (peaks of the H$\alpha>3$~\AA\ and H$\alpha<3$~\AA\ subsamples, respectively), locate two distinct populations when placed on the diagnostic diagrams. The high-EW set falls in the star-forming region of both BPT diagrams ($\log[\text{NII}]/\text{H}\alpha\approx-0.54$, $\log[\text{OIII}]/\text{H}\beta\approx-0.53$) and at young ages ($\log\text{Age}\approx9.1$). The low-EW set, in contrast, falls in the low-ionisation (AGN/DIG) region of the BPT diagrams ($\log[\text{NII}]/\text{H}\alpha\approx0.03$, $\log[\text{OIII}]/\text{H}\beta\approx0.21$) and at old ages ($\log\text{Age}\approx9.8$). The blue and red star symbols in the bottom panels of Figure~\ref{Fig_emmission} mark these two positions. This is consistent with the picture in which the high-EW component traces star-forming regions, while the low-EW component is associated with diffuse ionised gas (DIG), gas ionised by the old stellar population and the harder radiation field of evolved sources, in galaxies whose continua are dominated by old stars \citep{cidfernandes2010}.

\begin{table}[htbp]
\centering
\caption{Peak equivalent widths of the emission lines in the H$\alpha>3$~\AA\ (star-forming) and H$\alpha<3$~\AA\ (quiescent/DIG) subsamples. Peak positions are the KDE modes of the $\log_{10}$\,EW distributions of each subsample; for H$\alpha$ the two peaks are the modes of the two skew-normal components fitted to the full distribution.}
\label{tab:emission_peaks}
\begin{tabular}{lcc}
\toprule
\textbf{Line} & \textbf{EW peak (H$\alpha>3$)} & \textbf{EW peak (H$\alpha<3$)} \\
\midrule
H$\delta$       & 1.54 & 0.46  \\
H$\gamma$       & 2.50 & 0.46  \\
H$\beta$        & 4.93 & 0.38 \\
H$\alpha$       & 20.72 & 0.84  \\
\relax [OII]$\lambda$3726 & 4.99 & 0.86  \\
\relax [OII]$\lambda$3729 & 5.72 & 0.92  \\
\relax [SII]$\lambda$6716 & 3.78 & 0.43\\
\relax [SII]$\lambda$6731 & 2.71 & 0.28  \\
HeI$\lambda$5876 & 0.64 & 0.09  \\
\relax [OIII]$\lambda$5007 & 1.96 & 0.82  \\
\relax [OI]$\lambda$6300 & 1.18 & 0.18  \\
\relax [NII]$\lambda$6583 & 8.00 & 1.21  \\
\bottomrule
\end{tabular}
\end{table}

We note that the reliability of a given EW measurement decreases towards small values. At EW$\lesssim$1~\AA, a measured value can equally reflect a genuine weak emission component, for example the diffuse ionised gas traced by the low-EW peak, or a noise-driven overfit, in which random fluctuations are fit by an emission-line template and produce a spurious positive EW. As seen in Table~\ref{tab:emission_peaks}, even the low-EW peaks of the strongest lines fall in this regime ($\sim$0.1--1~\AA), where the noise floor is no longer negligible. We therefore caution the reader, and in particular users of our catalogue, that very low EW values should be interpreted with care: they may be real or spurious, and secure detections should be selected with an appropriate signal-to-noise or uncertainty cut, with the remaining weak values treated as upper limits where necessary.

\subsection{Kinematic results}
\label{sect:kinematic_results}

\begin{figure}
\centering
\includegraphics[width=15cm, angle=0]{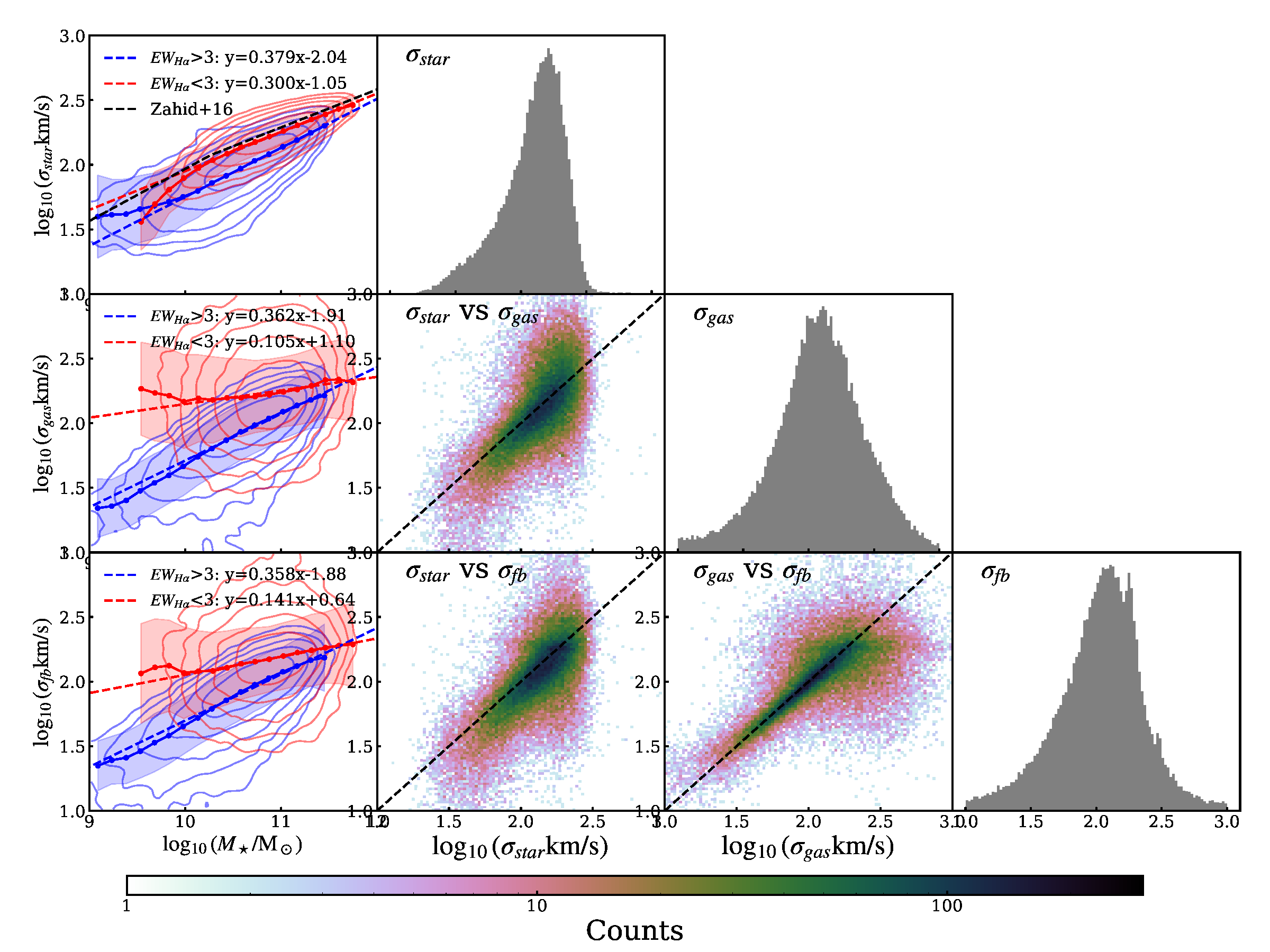}
\caption{Kinematic results from the full-spectrum fitting. Top: stellar velocity dispersion and gas velocity dispersion as functions of stellar mass, split by H$\alpha$ EW; the blue and red dashed lines mark the best-fit binned-median relations for the SF and QG populations, matching the slopes quoted in Table~\ref{tab:kinematics}, and the dashed black line shows the relation from \citet{zahid2016}. Middle and bottom panels: correlations between the three kinematic components ($\sigma_{\star}$, $\sigma_{\text{gas}}$, $\sigma_{\text{forbid}}$).}
\label{Fig4}
\end{figure}

A third key aspect is the kinematics of the stellar and gas components. Figure~\ref{Fig4} presents the kinematic results of our fitting. Following the discussion of the emission lines (Section~\ref{sect:emission_results}), we use the H$\alpha$ equivalent width to separate the two galaxy populations into star-forming (SF) and quiescent (QG) galaxies.

The right 3 column panels of Figure~\ref{Fig4} compare the velocity dispersions of the three kinematic components: the stellar component ($\sigma_{\star}$), the Balmer-line gas ($\sigma_{\mathrm{gas}}$), and the forbidden-line gas ($\sigma_{\mathrm{forbid}}$). For star-forming galaxies, the stellar velocity dispersion is systematically larger than the gas value by $\sim$0.09\,dex ($\sigma_{\star}/\sigma_{\mathrm{gas}}\approx1.2$), consistent with the nebular gas tracing a dynamically colder interstellar medium. The Balmer and forbidden-line components, in contrast, are in excellent agreement, with median dispersions differing by only $\sim$0.005\,dex and a high correlation ($r\approx0.88$), indicating that the two gas components trace the same ionised-gas kinematics. In quiescent galaxies the emission lines are too weak to constrain the gas kinematics reliably, so we do not interpret their gas velocity dispersions physically.

The velocity dispersion--stellar mass ($\sigma_{\star}$--$M_{\star}$) relation is a fundamental scaling relation of galaxies\citep{faber1989,donofrio2006,bernardi2003,cappellari2013}. The left panels of Figure~\ref{Fig4} show the velocity dispersion as a function of stellar mass for the two populations, and Table~\ref{tab:kinematics} lists the best-fit parameters of the linear relations. For star-forming galaxies, the three components follow nearly identical relations. This close agreement indicates that all three tracers respond to the same gravitational potential.

Quiescent galaxies have a larger stellar velocity dispersion than star-forming galaxies, and a shallower slope ($0.30$) that closely matches the high-mass relation of \citet{zahid2016} for SDSS galaxies ($0.29$). The gas velocity dispersions of quiescent galaxies, in contrast, increase only slowly with stellar mass. The binned medians underlying these trends are tightly constrained, but they should be interpreted with caution: the emission lines of quiescent galaxies are too weak for an accurate gas dispersion measurement, and the individual values carry large uncertainties. The shallow trends are likely contaminated by noise in the weak-line fits.

\begin{table}[htbp]
\centering
\caption{Parameters of the velocity dispersion--stellar mass relations for the stellar, Balmer, and forbidden-line components, separated by star formation activity. The slope and intercept are obtained by fitting the binned medians over $\log M_{\star}>10$, with $\log M_{\star}$ in units of $\mathrm{M_\odot}$. Quoted uncertainties are the $1\sigma$ bootstrap errors (200 resamplings of the parent galaxy sample) of the binned-median linear fits. The last row gives the high-mass ($\log M_{\star}>10.26$) relation of \citet{zahid2016} for comparison.}
\label{tab:kinematics}
\begin{tabular}{llrr}
\toprule
Component & Class & slope & intercept \\
\midrule
$\sigma_{\star}$                & SF & $0.379\pm0.005$ & $-2.04\pm0.06$ \\
$\sigma_{\star}$                & QG & $0.300\pm0.002$ & $-1.05\pm0.02$ \\
$\sigma_{\mathrm{gas}}$ (Balmer) & SF & $0.362\pm0.006$ & $-1.91\pm0.07$ \\
$\sigma_{\mathrm{gas}}$ (Balmer) & QG & $0.105\pm0.008$ & $1.10\pm0.09$ \\
$\sigma_{\mathrm{forbid}}$       & SF & $0.358\pm0.006$ & $-1.88\pm0.06$ \\
$\sigma_{\mathrm{forbid}}$       & QG & $0.141\pm0.007$ & $0.64\pm0.07$ \\
\midrule
\citet{zahid2016} ($\log M_{\star}>10.26$) & & 0.29 & $-0.93$ \\
\bottomrule
\end{tabular}%
\end{table}

We caution that the velocity dispersions presented here are aperture dispersions measured within the LAMOST fibres (3.3~arcsec diameter); the fraction of the galaxy sampled by the fibre varies with distance and galaxy size, so the scaling relations above may be affected by aperture effects \citep{zhu2023}, and a detailed aperture correction is beyond the scope of this catalogue paper. In addition, the velocity dispersions of low-mass galaxies are generally small, approaching the limits set by the spectral resolution and the fitting templates. Their measurements are sensitive to a number of systematic effects and may be subject to a lower boundary, so the low-mass end of the relations should be interpreted with caution.

\section{Summary}
\label{sect:summary}

The LAMOST Galaxy Parameter Catalog (LAGC) provides a dedicated, homogeneous set of stellar population, kinematic, and emission-line measurements of LAMOST galaxy spectra, and is demonstrated here with the pilot-release value-added catalogue, built from the full galaxy sample of the latest LAMOST data release without additional selection. Each spectrum is fitted with a library of 144 E-MILES simple stellar population templates and 19 Gaussian emission-line templates, organised into three kinematically independent components (stellar, Balmer, and forbidden lines). The resulting catalogue lists light-weighted and mass-weighted ages and metallicities, the six kinematic parameters of the stellar and gas components, the stellar mass-to-light ratio, and equivalent widths for all 19 emission lines, together with their bootstrap-based uncertainties. Alongside the integrated quantities, the catalogue also stores the full set of 144 SSP weights and the five-band signal-to-noise ratios, so that users can reconstruct or re-weight the stellar populations for their own purposes.

The SEW method and a wavelength-dependent line spread function are the two prerequisites for these measurements. SEW operates in equivalent-width space, which removes the multiplicative effects of flux calibration and dust attenuation and is therefore well suited to the LAMOST spectra, which are poorly flux calibrated. The line spread function is measured by fitting the lamp-line profiles predicted by the machine-learning lamp-spectrum model of \citet{liu2026}, whose wavelength-dependent predictions prevent an LSF mismatch from being absorbed into biased kinematics and stellar populations.

The reliability of the catalogue is established through both internal and external comparisons. Internally, repeated LAMOST observations of the same galaxies show median deviations consistent with zero at all signal-to-noise ratios, with a scatter that decreases as photon statistics improve, and the normalised residuals follow a standard normal distribution with unit variance, confirming that the bootstrap uncertainties are well calibrated. Externally, comparison with SDSS for 125,388 matched galaxies shows small offsets in the stellar population parameters and the stellar velocity dispersion, together with broad agreement with three earlier LAMOST catalogues; the systematic offset of $\sim$50~km\,s$^{-1}$ in the gas velocity dispersion is attributed to the different LSF modelling.

The measurements also provide a science-level validation of the catalogue by reproducing the well-established relations of the local galaxy population. Old stellar populations follow the stellar mass--metallicity relation. The H$\alpha$ equivalent-width distribution is bimodal, with a deep valley at $\sim$3~\AA\ separating star-forming galaxies from quiescent ones dominated by diffuse ionised gas; BPT diagrams built from equivalent-width ratios recover the familiar star-forming sequence, and the strong anti-correlation between the H$\alpha$ equivalent width and the light-weighted age confirms that these measurements trace recent star formation. For star-forming galaxies, the difference between the light-weighted and mass-weighted ages and metallicities is consistent with the downsizing and gas-inflow scenarios, indicating that the catalogue can trace the recent star formation and enrichment history across a large, homogeneous sample from the stellar side. The stellar, Balmer, and forbidden-line components of star-forming galaxies follow nearly identical velocity dispersion--stellar mass relations, indicating that they trace the same gravitational potential, whereas quiescent galaxies show a shallower relation consistent with SDSS. Reproducing these expected relations confirms that the catalogue is free of major systematic biases and provides a homogeneous basis for statistical studies of galaxy formation and evolution.

The pilot-release value-added catalogue is publicly available at the China-VO/NADC data platform, and the catalogue series is updated continuously as new LAMOST observations are processed, with the measurements of each subsequent release appended to the series. Alongside these updates, the template library will be extended in future releases: as described in Section~\ref{sect:templates}, the fiducial template set adopted for the pilot release will be supplemented with $\alpha$-enhanced and TP-AGB-rich stellar populations and non-solar abundance ratios.

Looking ahead, the methods developed for LAG2P provide a useful reference for the spectral processing of other fibre-fed surveys, which face similar flux-calibration and line-spread-function challenges. At the same time, the value-added catalogue is expected to lead to interesting scientific discoveries: its simultaneous measurements of stellar populations, kinematics, and emission lines open new avenues for statistical studies of galaxy formation and evolution.

\begin{acknowledgements}
This work is supported by the National Natural Science Foundation of China (NSFC) under No.~12403016.

Guoshoujing Telescope (the Large Sky Area Multi-Object Fiber Spectroscopic Telescope LAMOST) is a National Major Scientific Project built by the Chinese Academy of Sciences. Funding for the project has been provided by the National Development and Reform Commission. LAMOST is operated and managed by the National Astronomical Observatories, Chinese Academy of Sciences.
\end{acknowledgements}

\appendix
\section{Representative fitting examples}
\label{app:examples}

Representative full-spectrum fits of four galaxies with independent LAMOST repeat observations and matching SDSS spectra are shown in Figure~\ref{Fig3}, all with $r$-band SNR $> 15$. Despite the generally high SNR, the observed flux distributions (middle panels) show notable differences both between the two surveys and between repeat LAMOST exposures of the same galaxy, arising from time-varying instrumental and calibration effects: the first two galaxies exhibit stitching artifacts at the junction of the blue and red arms, while the latter two show broader continuum offsets. Once the data are transformed into the equivalent-width space (bottom panels), these continuum-level differences are effectively removed, and the EW spectra of the independent observations overlap within their error envelopes. The SSP weight maps (top panels) agree well between the two LAMOST observations, whereas the SDSS maps appear more scattered, because the SDSS spectra are fitted without bootstrap resampling for computational efficiency. These examples illustrate the advantage of the SEW approach over traditional additive-polynomial fitting, in which continuum artifacts such as stitching discontinuities or global calibration offsets cannot be fully captured by low-order polynomials, and even high-order polynomials tend to overfit the noise, biasing the inferred stellar populations and kinematics (see \citet{lu2025sew}).

\begin{figure}
  \centering
   \includegraphics[width=150mm]{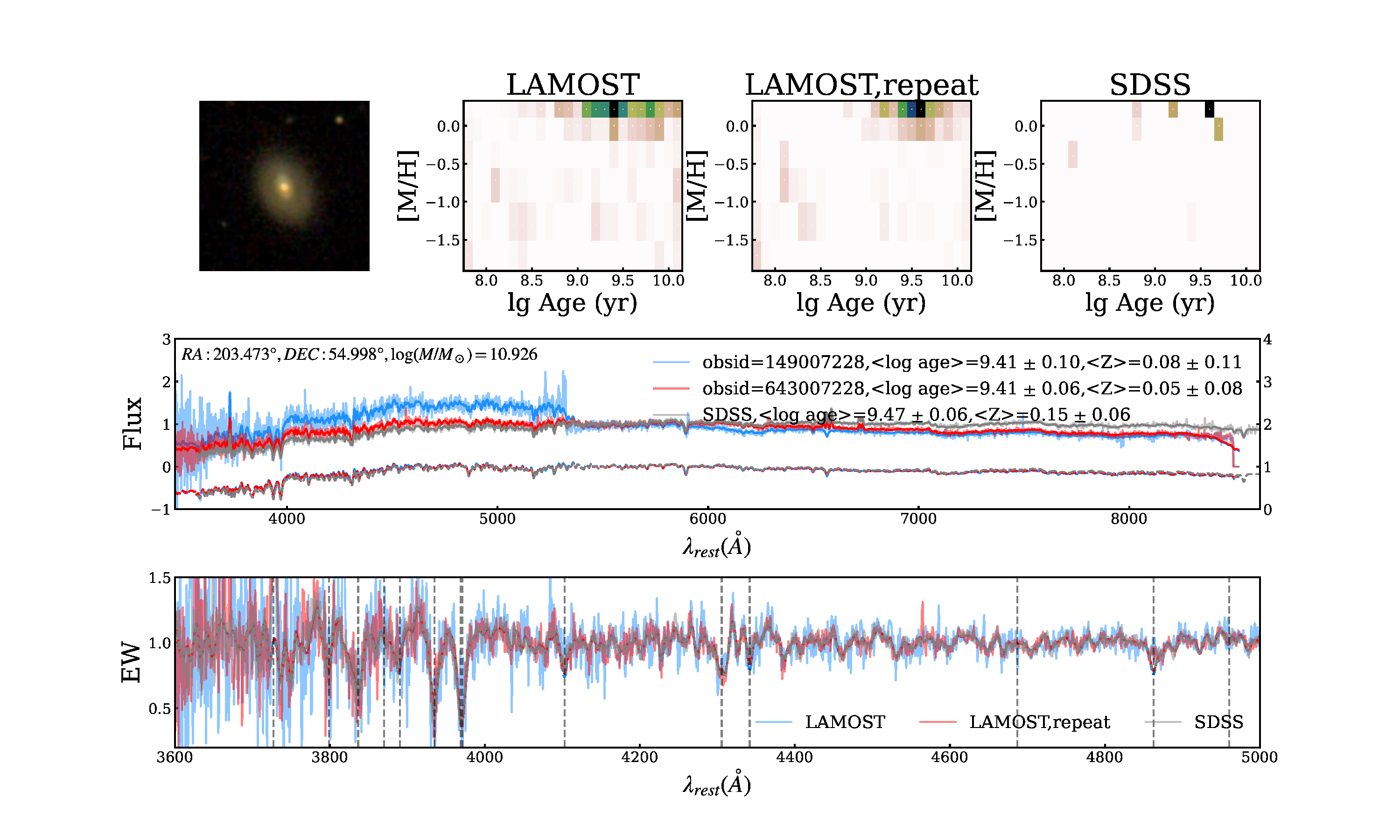}\\
   \includegraphics[width=150mm]{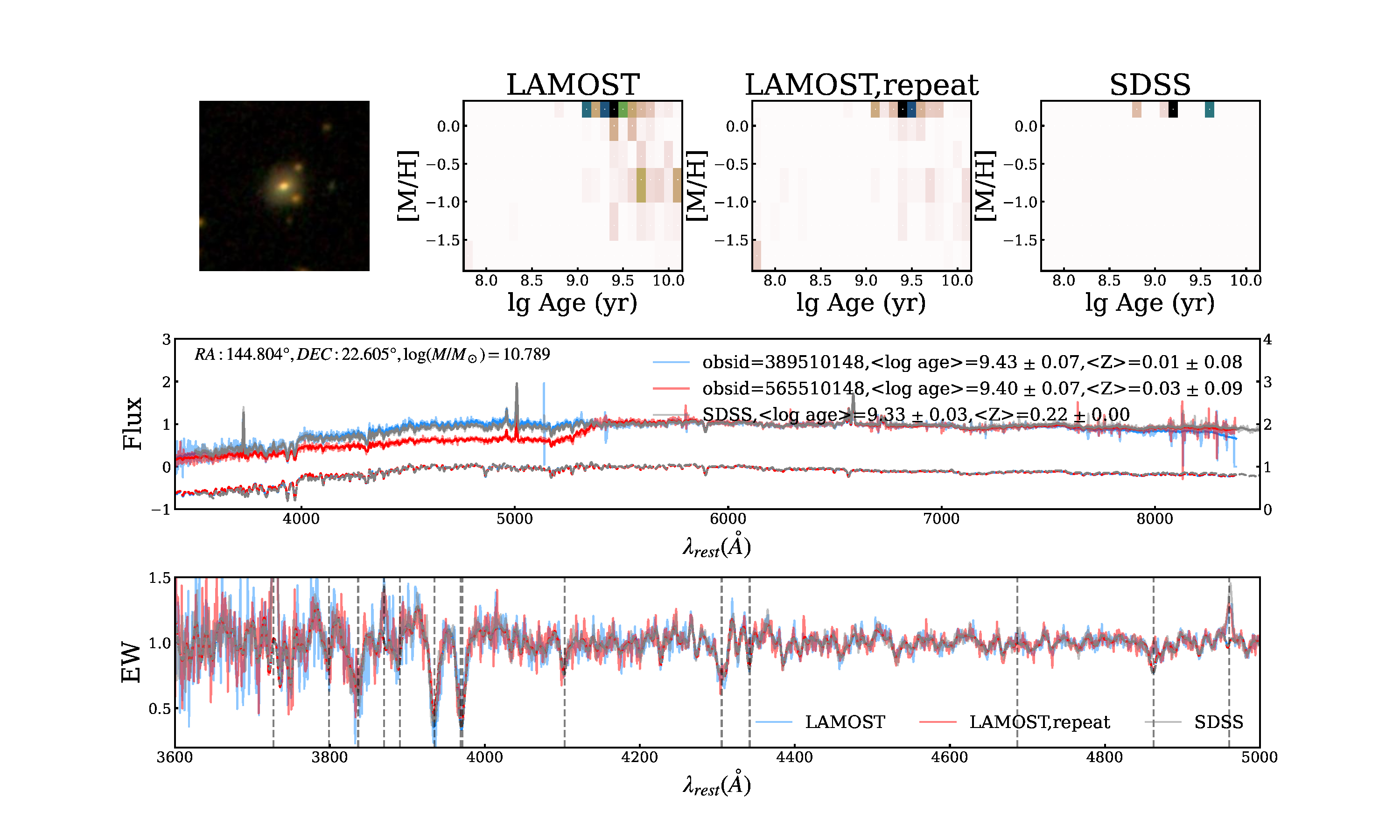}
\caption{Examples of full-spectrum fits for LAMOST galaxies with repeated LAMOST observations and matching SDSS spectra. For each galaxy, the top panels show the SDSS image and stellar population weight maps from each observation. Middle panels show the full spectral fits (observed flux in light curves, best-fit models in solid lines), and bottom panels show the equivalent-width spectra. Blue and red curves represent LAMOST first and repeat observations; black curves represent SDSS fits.}
\label{Fig3}
\end{figure}

\addtocounter{figure}{-1}
\begin{figure}
  \centering
   \includegraphics[width=150mm]{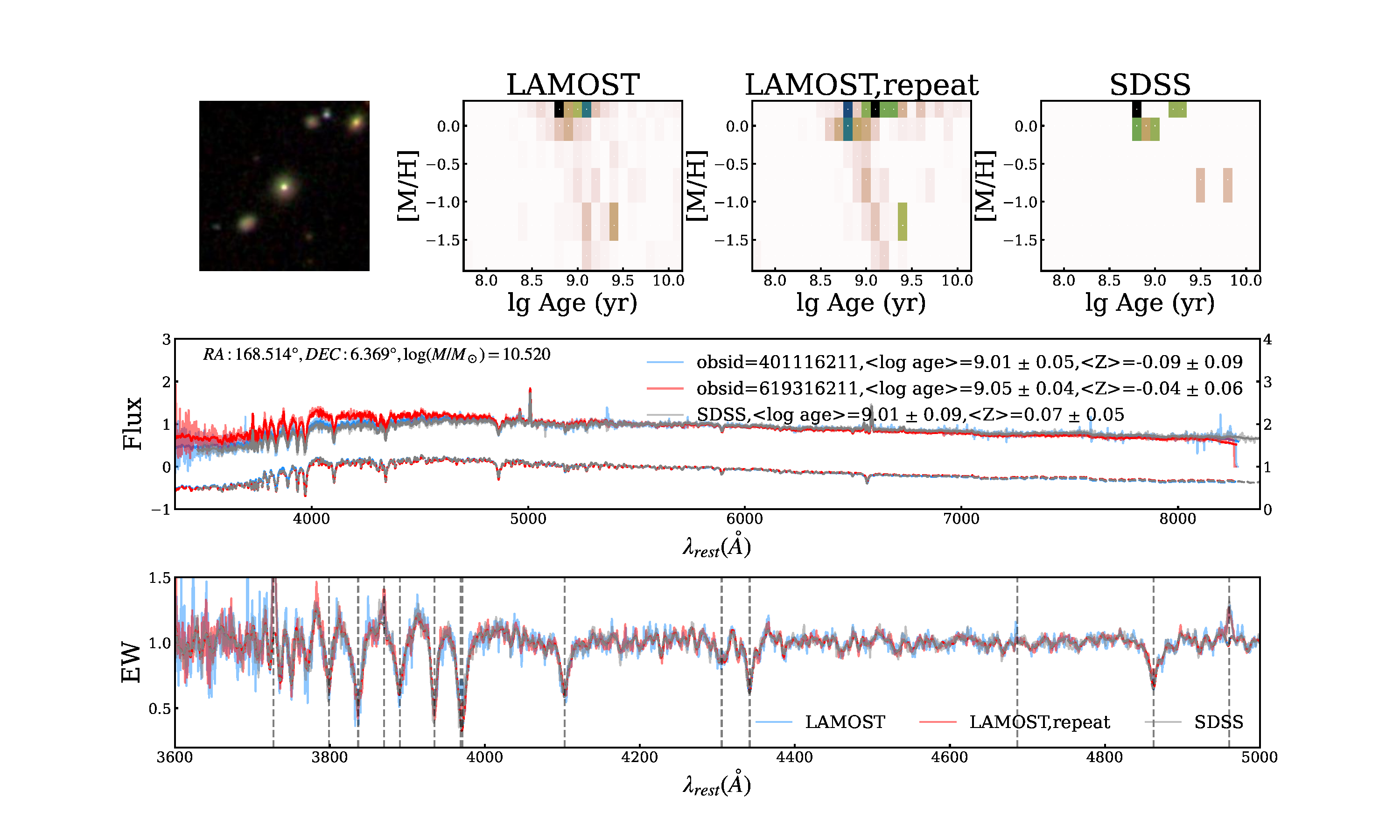}\\
   \includegraphics[width=150mm]{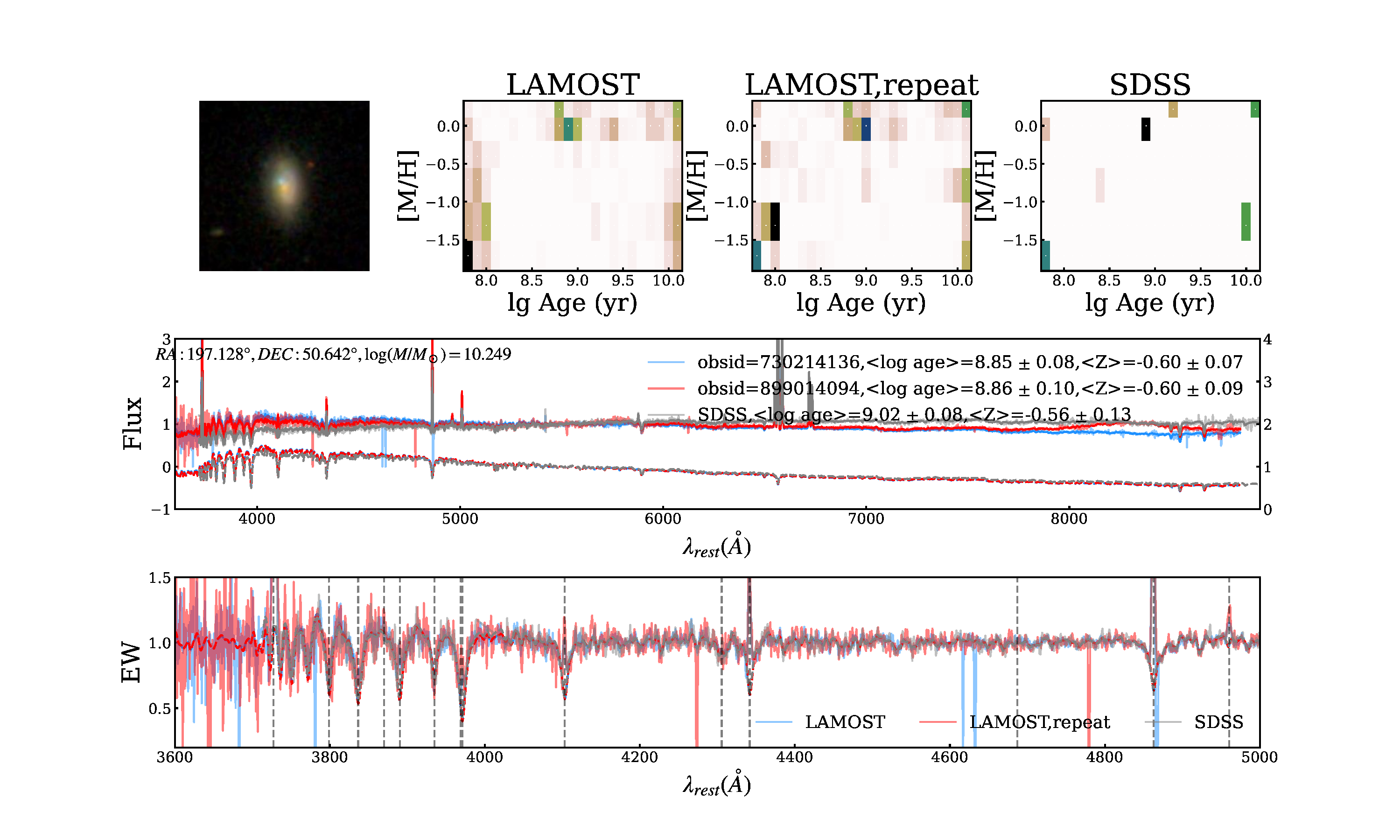}
\caption[]{(continued)}
\end{figure}

\section{Comparison with previous LAMOST catalogues}
\label{app:previous}

As an additional external validation, we compare our catalogue with three previous LAMOST-based measurements: the stellar velocity dispersion catalogue of \citet{napolitano2020}, the emission-line measurements of \citet{wang2018}, and the stellar population catalogue of \citet{wang2022}. These catalogues were derived with independent methods and template libraries; they allow us to assess the systematic consistency of our measurements and to quantify the impact of methodological choices.

\subsection{Stellar velocity dispersion}

\begin{figure}
\centering
\includegraphics[width=15cm, angle=0]{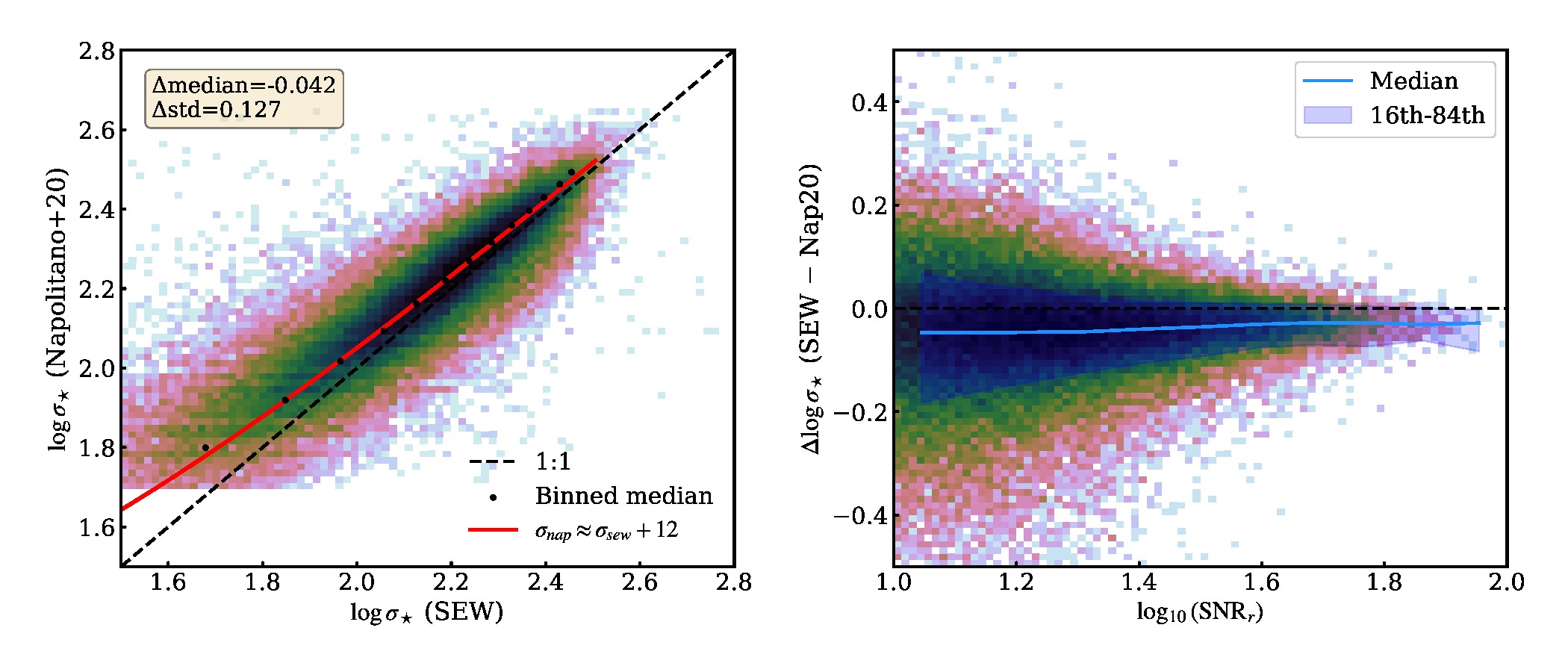}
\caption{Comparison of stellar velocity dispersions between this work (SEW) and the LAMOST DR7 catalogue of \citet{napolitano2020}. Left: direct comparison of $\log\sigma_{\star}$; the dashed line marks the one-to-one relation. Right: $\Delta\log\sigma_{\star}$ as a function of SNR, with the median and 16th--84th percentile ranges.}
\label{Fig_nap}
\end{figure}

\citet{napolitano2020} measured central stellar velocity dispersions for $\sim$\,86,000 LAMOST DR7 galaxies using \textsc{pPXF} \citep{cappellari2017} with MILES templates and additive polynomial corrections for continuum variations and template mismatch. Figure~\ref{Fig_nap} compares the two measurements. The two catalogues agree well overall, with a median offset of $\Delta\log\sigma_{\star} = -0.04$\,dex and a scatter of 0.13\,dex. A closer inspection, however, reveals a small but systematic constant offset: the velocity dispersions of \citet{napolitano2020} are higher than ours by a constant $\sim$\,12~\text{km\,s}$^{-1}$.

This constant offset most likely reflects residual continuum and template mismatch in the previous analysis, as the additive polynomial corrections used by \citet{napolitano2020} are not ideally suited to the predominantly multiplicative flux calibration uncertainties of LAMOST spectra \citep{luo2015,wang2018}. The SEW method, by contrast, removes the continuum by construction and is largely immune to such errors. Combined with the well-calibrated bootstrap uncertainties validated by the normalised residuals from repeat observations (Figure~\ref{Fig_noise}), this suggests that the $\sim$\,12~\text{km\,s}$^{-1}$ offset reflects a bias in the previous analysis and that our $\sigma_{\star}$ measurements are more reliable.

\subsection{Emission lines}

\begin{figure}
\centering
\includegraphics[width=15cm, angle=0]{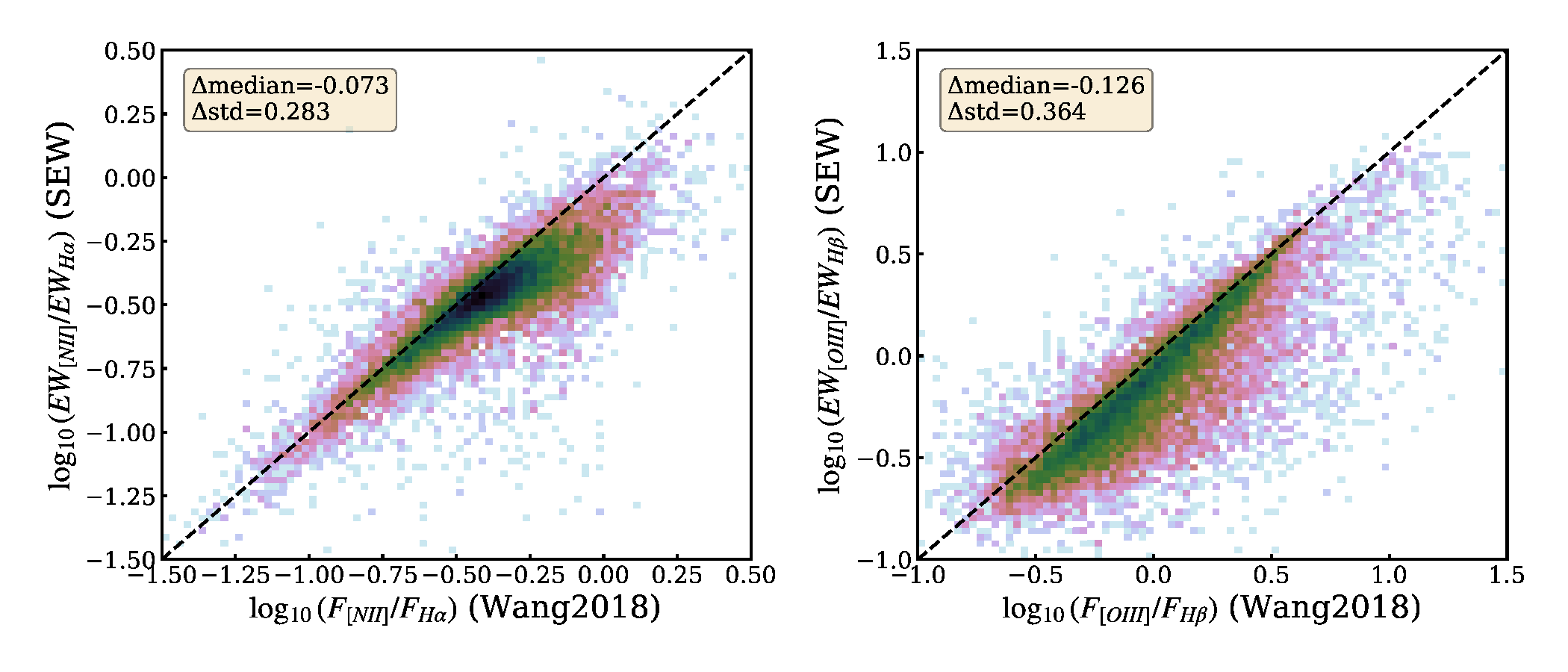}
\caption{Comparison of emission-line strength ratios between this work (equivalent-width ratios) and \citet{wang2018} (flux ratios). Left: $[\text{NII}]/\text{H}\alpha$; right: $[\text{OIII}]/\text{H}\beta$. The dashed lines mark the one-to-one relations.}
\label{Fig_emm}
\end{figure}

\citet{wang2018} provided emission-line fluxes for LAMOST DR4 galaxies. Since the SEW method measures equivalent widths rather than fluxes, we compare the line-strength ratios $[\text{NII}]/\text{H}\alpha$ and $[\text{OIII}]/\text{H}\beta$ rather than the absolute line strengths. Because the two lines in each pair have similar rest wavelengths, the continua underlying them are very close, so that the ratios are largely insensitive to the continuum normalisation and can be directly compared between equivalent-width and flux measurements. Figure~\ref{Fig_emm} shows the comparison. The two line ratios agree well overall, with scatters of 0.28 and 0.37\,dex for $[\text{NII}]/\text{H}\alpha$ and $[\text{OIII}]/\text{H}\beta$, respectively, but the measurements of \citet{wang2018} are systematically higher than ours by $\sim$\,0.07 and $\sim$\,0.13\,dex.

This offset most likely reflects, at least in part, differences in the stellar absorption-line subtraction between the two methods. These algorithmic differences, which are largely driven by the adopted LSF models and stellar template libraries, lead to residual differences in the subtracted absorption. Together with small differences between the continua underlying the two lines of each pair, they ultimately result in the relatively lower line ratios of our measurements.

\subsection{Stellar populations}

\begin{figure}
\centering
\includegraphics[width=15cm, angle=0]{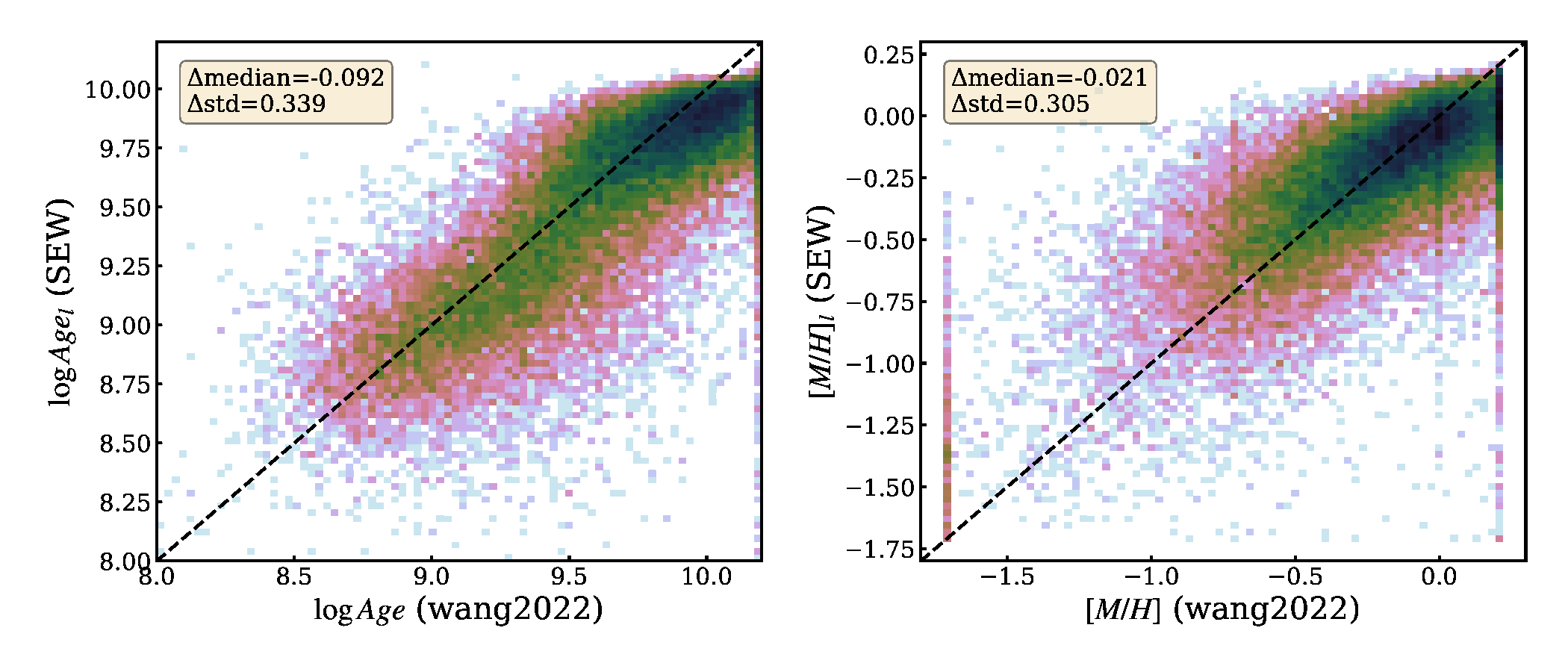}
\caption{Comparison of light-weighted stellar population parameters between this work (SEW) and \citet{wang2022}: light-weighted age (left) and metallicity (right). The dashed lines mark the one-to-one relations.}
\label{Fig_pop}
\end{figure}

\citet{wang2022} derived light-weighted stellar population parameters for $\sim$\,43,000 LAMOST DR7 galaxies using a fitting procedure based on small-scale spectral features to mitigate flux calibration uncertainties. Figure~\ref{Fig_pop} compares the light-weighted ages and metallicities. The two catalogues are in good agreement, with median offsets of $-0.09$\,dex in $\log\text{Age}$ and $-0.02$\,dex in $[\text{M/H}]$ and scatters of 0.34 and 0.30\,dex, respectively. Our mass-weighted ages and metallicities are correspondingly higher than theirs, by $+0.18$ and $+0.08$\,dex, as expected from the different weighting of the young and old stellar components. These small offsets are most likely driven by differences in the LSF models and the stellar population synthesis procedures (e.g., template libraries and fitting algorithms) between the two works. Although slight differences remain, the overall consistency is satisfactory. In contrast to the previous catalogue, our measurements come with self-consistent bootstrap-based uncertainty estimates, making them more reliable for statistical analyses.

In summary, our measurements are broadly consistent with these three independent LAMOST-based catalogues. The remaining systematic differences, namely the $\sim$\,12~\text{km\,s}$^{-1}$ offset in $\sigma_{\star}$ relative to \citet{napolitano2020} and the higher line ratios of \citet{wang2018}, can be understood as direct consequences of the different treatments of the continuum and of stellar absorption in the respective methods, further supporting the reliability of the SEW approach for large spectroscopic surveys.

\end{document}